\documentstyle[manuscript,aps,tighten,graphicx]{revtex}
\begin{document}
\draft
\title{$\bbox{^3P_2}$-$\bbox{^3F_2}$ pairing in neutron matter\\
       with modern nucleon-nucleon potentials}
\author{M. Baldo$^1$, \O. Elgar\o y$^2$, L. Engvik$^2$, 
 M. Hjorth-Jensen$^3$, and H.-J. Schulze$^1$}
\address{$^1$ Sezione INFN, Universit\`{a} di Catania, Corso Italia 57,
         I-95129 Catania, Italy}
\address{$^2$Department of Physics, University of Oslo, N-0316 Oslo, Norway}
\address{$^3$Nordita, Blegdamsvej 17, DK-2100 Copenhagen \O, Denmark}
\maketitle
\date{}
\begin{abstract}
  We present results for the $^3P_2$-$^3F_2$ pairing gap in 
  neutron matter with several realistic nucleon-nucleon potentials,
  in particular with recent, phase-shift equivalent potentials.   
  We find that their predictions for the gap cannot 
  be trusted at densities above $\rho\approx 1.7\rho_0$, where 
  $\rho_0$ is the saturation density for symmetric nuclear matter.      
  In order to make predictions above that density, potential 
  models which fit the nucleon-nucleon phase shifts up 
  to about 1 GeV are required.  
\end{abstract}

\section{Introduction}

The presence of neutron superfluidity in the crust and the inner part 
of neutron stars is one of the features that are considered well established 
in the physics of these compact stellar objects. 
At low density, and therefore
in the outer part of a neutron star, the neutron superfluidity should be
mainly in the $^1S_0$ channel. 
At higher density, the nuclei in the crust dissolve, and one 
expects a region consisting of a quantum liquid of neutrons and 
protons in beta equilibrium. 
The proton contaminant should be superfluid 
in the $^1S_0$ channel, while neutron superfluidity is expected to  
occur mainly in the coupled $^3P_2$-$^3F_2$ two-neutron channel. 
In the core of the star any superfluid phase should finally disappear.
 
The presence of two different superfluid regimes 
is suggested by the known trend of the nucleon-nucleon (NN) phase shifts 
in each scattering channel. 
In both the $^1S_0$ and $^3P_2$-$^3F_2$ channels the
phase shifts indicate that the NN interaction is attractive. 
In particular for the $^1S_0$ channel, the occurrence of 
the well known virtual state in the neutron-neutron channel
strongly suggests the possibility of a pairing condensate at low density, 
while for the $^3P_2$-$^3F_2$ channel the interaction becomes strongly attractive only
at higher energy, which therefore suggests a possible pairing condensate
in this channel at higher densities. In recent years the BCS gap equation
has actually been solved with realistic interactions, and the results confirm
these expectations. 

The $^1S_0$ neutron superfluid is relevant for phenomena
that can occur in the inner crust of neutron stars, like the 
formation of glitches, which seem to be related to vortex pinning  
of the superfluid phase in the solid crust \cite{glitch}. 
The results of different groups are in close agreement
on the $^1S_0$ pairing gap values and on its density dependence, which
shows a peak value of about 3 MeV at a Fermi momentum close to
$k_F \approx 0.8\; {\rm fm}^{-1}$ \cite{bcll90,kkc96,eh98,sclbl96}. 
All these calculations adopt the bare
NN interaction as the pairing force, and it has been pointed out
that the screening by the medium of the interaction could strongly reduce
the pairing strength in this channel \cite{sclbl96,chen86,ains89}. 
However, the issue of the 
many-body calculation of the pairing effective interaction is a complex
one and still far from a satisfactory solution.

The precise knowledge of the $^3P_2$-$^3F_2$ pairing gap is of 
paramount relevance for, e.g., the cooling of neutron stars, 
and different values correspond to drastically
different scenarios for the cooling process \cite{nstar}.
Unfortunately, only few and partly
contradictory calculations of this quantity exist in the literature, 
even at the level of the bare NN interaction 
\cite{amu85,bcll92,taka93,elga96,khodel97}. 
However, when comparing the results, one should note that the  
NN potentials used in these calculations are not phase-shift 
equivalent, i.e., they do not 
predict exactly the same NN phase shifts.  
Furthermore, for the interactions used in 
Refs.~\cite{amu85,bcll92,taka93,elga96} the predicted 
phase shifts do not agree accurately with modern phase shift 
analyses, and the fit of the NN data has typically 
$\chi^2/{\rm datum}\approx 3$.  During the last years, progress has 
been made not only in the accuracy and the consistency of the 
phase-shift analysis, but also in the fit of realistic NN potentials 
to these data.  As a result, several new NN potentials have 
been constructed which fit the world data for $pp$ and $np$ scattering 
below 350 MeV with high precision.  Potentials like the recent 
Argonne $V_{18}$ \cite{v18}, the CD-Bonn \cite{bon} 
or the new Nijmegen potentials \cite{nij} yield a 
$\chi^2/{\rm datum}$ of about 1 and may be called phase-shift 
equivalent.  

Our aim in this paper is to compare the predictions of 
the new potentials for the $^3P_2$-$^3F_2$ gap in neutron matter.  
We will also, for the sake of completeness, include results 
with some of the ``old'' interactions, namely the Paris \cite{par}, 
Argonne $V_{14}$ \cite{v14}, and Bonn B \cite{mach89} potentials.  
The main focus will, however, be on the new, phase-shift  
equivalent potentials, and whether the improved accuracy in 
the fits to the NN scattering data leads to better agreement 
in the predictions for the $^3P_2$-$^3F_2$ energy gap.  
If differences are found, we try to trace them back to 
features of the NN potentials.  
To be able to do so, we will keep the many-body formalism as 
simple as possible.  First of all, we will use   
the bare NN interaction as kernel in the gap equations, and 
thus neglect higher-order contributions from, e.g., medium 
polarization effects.  The in-medium single-particle energies 
will be calculated in the Brueckner-Hartree-Fock (BHF) approximation, 
but we will also use free single-particle energies, because this 
makes the comparison of the results with the various potentials 
more transparent, since any differences are then solely due to 
differences in the $^3P_2$-$^3F_2$ wave of the potentials.  
We think it is useful to try to understand the results at the 
simplest level of many-body theory before proceeding to include 
more complicated effects in the description of $^3P_2$-$^3F_2$ pairing.   
As we will demonstrate, progress in the construction 
of NN interactions is necessary before the $^3P_2$-$^3F_2$ energy gap can be 
calculated reliably from microscopic many-body theory.   

This work falls in six sections. 
The equations for solving the pairing gap are briefly reviewed 
in the next section, while in Section \ref{sec:num_methods} 
we discuss the reliability of various numerical approaches to the 
solution of the pairing gap. Features of the various nucleon-nucleon
interaction models employed are presented in Section \ref{sec:nn_pots},
while our results for the pairing gap with these potentials
are discussed in Section \ref{sec:results}. Finally, we summarize
our findings in Section \ref{sec:conclusions}.   

\section{Gap equation for the $\bbox{^3P_2}$-$\bbox{^3F_2}$ channel}

The gap equation for pairing in non-isotropic partial waves is in general
more complex than in the simplest $s$-wave case, in particular in neutron and 
nuclear matter, where the tensor interaction can couple two 
different partial waves \cite{taka93,bls95}. 
This is indeed the situation for the $^3P_2$-$^3F_2$ neutron channel. 
In order to achieve a simplified, yet accurate, numerical treatment, 
we use in this work the angle average approximation 
explained in this section.

For the sake of a clear presentation, we disregard for the moment the
spin degrees of freedom and the tensor interaction.
Then, we start with the Gorkov equations \cite{gorkov}, 
which involve the propagator
$G(\bbox{k},\omega)$, the anomalous propagator $F(\bbox{k},\omega)$, and the
gap function $\Delta(\bbox{k})$: 
\begin{eqnarray}
 \left( \begin{array}{rr} \omega-\epsilon(\bbox{k}) & -\Delta(\bbox{k}) \\ 
  -\Delta^\dagger(\bbox{k}) & \omega+\epsilon(\bbox{k})
 \end{array} \right)
 \left(\begin{array}{c} G \\ F^\dagger \end{array}\right)(\bbox{k},\omega) =
 \left( \begin{array}{c} 1 \\ 0 \end{array} \right) \:,
\end{eqnarray}
where $\epsilon(\bbox{k}) = e(\bbox{k}) - \mu$, $\mu$ being the 
chemical potential and $e(\bbox{k})$ the single-particle spectrum. 
The quasi-particle energy $E(\bbox{k})$ is the solution of the corresponding
secular equation and is given by
\begin{equation}
  E(\bbox{k})^2 = \epsilon(\bbox{k})^2 + |\Delta(\bbox{k})|^2 \:.
\end{equation}
The anisotropic gap function $\Delta(\bbox{k})$ is to be determined from the gap
equation
\begin{equation}
  \Delta(\bbox{k}) = - \sum_{\bbox{k}'} \langle \bbox{k} | V | \bbox{k}' 
\rangle
  {\Delta(\bbox{k}')\over 2E(\bbox{k}') } \:.
\end{equation}
The angle-dependent energy denominator in this equation prevents 
a straightforward separation into the different partial wave components
by expanding the potential,
\begin{equation}
  \langle \bbox{k} | V | \bbox{k}' \rangle = 
  4\pi \sum_l (2l+1) P_l(\bbox{\hat{k}\cdot\hat{k}'}) V_l(k,k') \:,
\end{equation}
and the gap function,
\begin{equation}
  \Delta(\bbox{k}) = 
  \sum_{l,m} \sqrt{4\pi\over 2l+1} Y_{lm}(\bbox{\hat{k}}) \Delta_{lm}(k) \:.
\end{equation}
However, after performing an angle average approximation for the gap in the
quasi-particle energy,
\begin{equation}
  |\Delta(\bbox{k})|^2 \rightarrow D(k)^2 \equiv 
  {1\over 4\pi} \int d\bbox{\hat{k}}\, |\Delta(\bbox{k})|^2 =
  \sum_{l,m} {1\over 2l+1} |\Delta_{lm}(k)|^2 \:,
\end{equation}
the kernels of the coupled integral equations become isotropic, and one can 
see that the different $m$-components become uncoupled and all equal. 
One obtains the following equations for the 
partial wave components of the gap function:
\begin{equation}
  \Delta_l(k) = - {1\over\pi} \int_0^\infty 
  {V_l(k,k') \over \sqrt{ \epsilon(k')^2 + 
  \left[\sum_{l'} \Delta_{l'}(k')^2 \right] }}
  \Delta_l(k') \:.
\end{equation}
Note that there is no dependence on the quantum number $m$ in these 
equations, however, they still couple the components of the
gap function with different $l$ 
($^1S_0$, $^3P_0$, $^3P_1$, $^3P_2$, $^1D_2$, $^3F_2$, etc.~in neutron matter) 
via the energy denominator.
Fortunately, in practice the different components $V_l$ of the potential
act mainly in non-overlapping intervals in density, 
and therefore also this coupling can usually be disregarded.

The addition of spin degrees of freedom and of the tensor force does not
change the picture qualitatively, and is explained in detail 
in Refs.~\cite{taka93,bls95}.
The only modification is the introduction 
of an additional $2\times2$ matrix structure due to the tensor coupling
of the $^3P_2$ and $^3F_2$ channels: 
\begin{mathletters}
\begin{eqnarray}
 \left( \begin{array}{c} \Delta_1 \\ \Delta_3 \end{array} \right)(k) &=&
 - {1\over\pi} \int_0^\infty dk' k'^2 {1\over E(k')}
 \left( \begin{array}{rr}
  V_{11} & -V_{13} \\ -V_{31} & V_{33}
 \end{array} \right)(k,k')
 \left(\begin{array}{c} \Delta_1 \\ \Delta_3 \end{array}\right)(k') \:,
\\
  E(k)^2 &=& [e(k)-e(k_F)]^2 + D(k)^2 \:,
\\
  D(k)^2 &=& \Delta_1(k)^2 + \Delta_3(k)^2 \:.
\label{e:gap2c}
\end{eqnarray}
\label{e:gap2}
\end{mathletters}
Here $e(k)=k^2/2m + U(k)$ are the single-particle energies, 
as obtained from a Brueckner-Hartree-Fock calculation, 
where $U(k)$ is the single-particle potential, calculated 
within the ``continuous choice'' scheme \cite{jjm76}.
The quantities 
\begin{equation}
  V_{ll'}(k,k') =  \int_0^\infty dr r^2 j_{l'}(k'r) V_{ll'}(r) j_l(kr)
\label{e:v}
\end{equation}
are the matrix elements of the bare interaction in 
the different coupled channels 
$(T=1;\;S=1;\,J=2;\,l,l'=1,3)$. 

It has been shown that the angle average approximation is an 
excellent approximation to the true solution that involves a gap function 
with ten components \cite{taka93,khodel97}, as long as one is 
only interested in the average value of the gap at the Fermi surface, 
$\Delta_F\equiv D(k_F)$, and not the angular dependence of the gap functions 
$\Delta_1(\bbox{k})$ and $\Delta_3(\bbox{k})$.  

\section{Numerical solution}
\label{sec:num_methods} 

The solution of the system of equations (\ref{e:gap2}) is numerically
not trivial, especially if the gap turns out to be much
smaller than the Fermi energy. This is because of the well known
logarithmic singularity of the BCS equation in the limit of zero pairing gap. 
In order to control more closely the numerical accuracy, we used in fact
three different methods:

One method is similar to the one described in Ref.~\cite{bcll90}.
We first obtain a separable form of the interaction. 
Since we need a high accuracy, we directly diagonalize the interaction
$V_{ll'}(k,k')$, taken in a discrete grid of momenta $\{k_i\}$, and then
we choose the first $n$ eigenvalues $\lambda_m$ with the largest
moduli, and the corresponding eigenvectors $v_m$.
One can then write
\begin{equation}
  V_{ll'}(k_i,k_j) \approx \sum_{m = 1}^n v_m(k_i) \lambda_m v_m(k_j) \:.
\end{equation}
The gap function can then also be expanded in the same eigenvectors, and
the original equations reduce to a set of $2n$ algebraic equations. 
The latter can be solved for the coefficients of the expansion
by iteration, following the scheme described in Ref.~\cite{bcll90}.
The rank $n$ of the separable form is increased
until a high degree of convergence is reached. One advantage
of the method is the possibility of using a very fine momentum grid,
since the algebraic equations are obtained by numerical integrations,
for which extremely accurate interpolation methods can be used.
In general, the grid points must be particularly dense in the interval
around the Fermi momentum, since there the kernel displays
an extremely narrow peak due to the small value of the pairing gap.
Furthermore, in general, the convergence in the rank $n$ is fast enough,
and therefore the number of coupled equations is never very large.
However, the accuracy in the diagonalization procedure is decreasing
with the rank of the matrix and it is difficult to have a precise 
estimate of the error.

In the second method \cite{krotsch} one starts by solving the gap equation
for the case of a constant pairing gap $\overline{\Delta}$ in the 
denominator. In a discrete momentum grid, this is equivalent
to an eigenvalue problem, namely to find the value of $\overline{\Delta}$
for which the kernel of the gap equation has eigenvalue one. 
The corresponding eigenvector is a first estimate of the gap function,
with the normalization $\Delta(k_F) = \overline{\Delta}$.
It is then inserted in the kernel to solve for the next estimate of 
$\overline{\Delta}$. 
In practice this method converges extremely fast (after a few iterations)
to the final solution.
The advantage of the method is that the original interaction is
used, without resorting to a separable form. 

The third method is to solve the coupled $^3P_2$-$^3F_2$ gap 
equations straightforwardly by iteration, starting from some 
suitable initial approximation to the functions $\Delta_1(k)$ and 
$\Delta_3(k)$.  
Also in this method, the interaction is used in its original form.  
If the interaction has a strong 
repulsive core, as is the case in the $^1S_0$ channel, this 
method can be difficult or even impossible to implement.  
However, the $^3P_2$-$^3F_2$ interaction is relatively weak, 
and the iteration scheme works well in this channel,  
provided that a fine momentum grid is used around the Fermi momentum.      
Details of the numerical implementation of this method are 
given in Ref.~\cite{elga96}.  

The comparison of the results obtained with the three methods was
quite rewarding. The numerical values of the gap functions were
in excellent agreement and hardly distinguishable
in all figures presented here. Therefore, in discussing the results we will 
not specify the method by which they were obtained. We believe that
the agreement between the three methods gives enough confidence
in the numerical precision of the results.

\section{The NN interactions}
\label{sec:nn_pots} 

Before discussing the solutions of the coupled $^3P_2$-$^3F_2$
gap equations, we give a short description of the models 
for the NN interaction employed in this paper.  

The older models, Paris, Argonne $V_{14}$, and Bonn B are 
described in detail in Refs.~\cite{par,v14,mach89}.  
They all have a $\chi^2/{\rm datum}$ in the range 2-3.  
The Argonne $V_{14}$ potential is a non-relativistic, purely local potential.  
The Paris potential incorporates explicit 
$\pi$-, $2\pi$-, and $\omega$-exchange.  
For the short-range part a phenomenological approach is used.  
The final potential is parameterized in terms of local Yukawa functions.  
The Bonn B potential is a one-boson-exchange (OBE) interaction, 
defined by the parameters of Table A.1 of Ref.~\cite{mach89}.  

The ``phase-shift equivalent'' potentials we will employ here 
are the recent models of the Nijmegen group \cite{nij}, 
the Argonne $V_{18}$ \cite{v18} potential and the charge-dependent Bonn 
potential (CD-Bonn) \cite{bon}.  
In 1993, the Nijmegen group presented a phase-shift analysis of all 
proton-proton and neutron-proton scattering data below 350 MeV with a 
$\chi^2/{\rm datum}$ of 0.99 for 4301 data entries \cite{nijpwa}.  
Fitted to this phase-shift analysis, the CD-Bonn potential 
has a $\chi^2/{\rm datum}$ of 1.03 and the same is true for 
the Nijm-I and Nijm-II potentials of the Nijmegen group \cite{nij}.  
The Argonne $V_{18}$ potential has a $\chi^2/{\rm datum}$ of 1.09.  

All these models are charge-dependent.  
Argonne $V_{18}$ and Nijm-II   
are non-relativistic potential models defined in terms of local  
functions, which are attached to various (non-relativistic) 
operators constructed from the spin, isospin and angular momentum 
operators of the interacting pair of nucleons.  
Such approaches to the NN potential have traditionally been quite 
popular since they are numerically easy to use in configuration space 
calculations. 
The Nijm-I model is similar to the Nijm-II model, but it includes 
also a momentum dependent term, see Eq.~(13) of Ref.~\cite{nij}, which 
may be interpreted as a non-local contribution to the central force.  
The CD-Bonn potential is based on the relativistic 
meson-exchange model of Ref.~\cite{mach89} which is 
non-local and cannot be described correctly in terms of local 
potential functions.
Instead, it is represented most conveniently in terms of partial waves.

Thus, the mathematical structure of the modern potentials is quite different,
although they all predict almost identical phase shifts
within their range of validity. 
This means that even though the potentials by construction 
give the same results on-shell, their behavior off the 
energy shell may be quite different.   
The implications of these differences for the symmetry energy 
of nuclear matter were discussed in Ref.~\cite{engv97}.   

\begin{figure}
\includegraphics[totalheight=11.cm,angle=0,bb=-50 170 350 720]{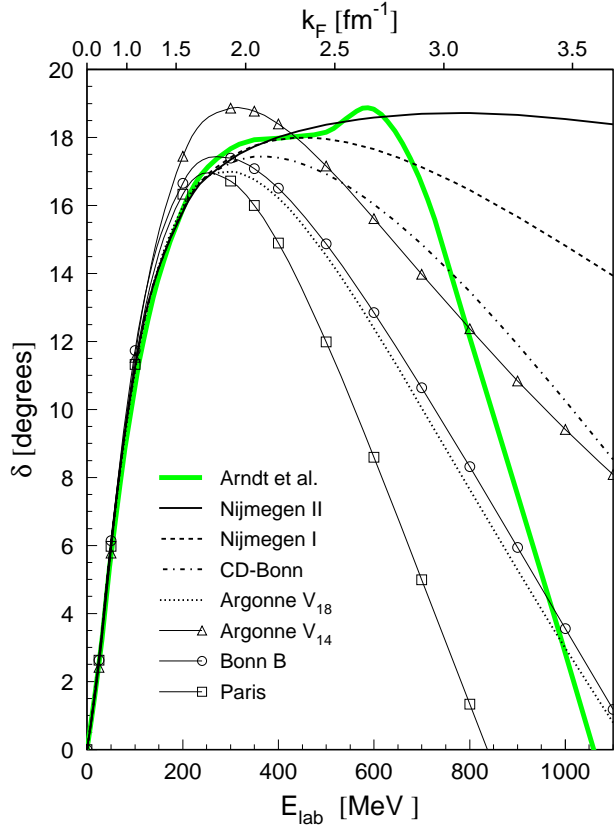}
\caption{$^3P_2$ phase shift predictions of different potentials 
         up to $E_{\rm lab}=1.1\;{\rm GeV}$, compared with the 
         phase shift analysis of Arndt et al.~\protect\cite{arndt97}.
         The ``old'' potentials are denoted by different symbols; 
         the ``modern'' potentials by different line styles.}
\label{fig:phaseshift}
\end{figure}

In order to illustrate the statements made above, 
and for a better understanding of the forthcoming results 
for the pairing gaps, 
we show in Fig.~\ref{fig:phaseshift} the predictions of the 
various potentials for the phase shifts in the $^3P_2$ ($T=1$) channel.
They have been calculated by solving the Lippmann-Schwinger 
equation as explained in Ref.~\cite{mach87}.
The figure shows predictions up to $E_{\rm lab}=1.1\;{\rm GeV}$,
but clearly scattering energies above $E_{\rm lab}=350\;{\rm MeV}$ 
amount to uncontrolled extrapolations beyond the intended range of validity
of the potential models, that have been fitted to scattering data
below $350\;{\rm MeV}$ only. 
The plot displays also a scale of equivalent Fermi momenta 
according to the relation
$E_{\rm lab}=(2k_F)^2/2m$ in order to facilitate the comparison
with the pairing gaps presented later.  The reader can see 
that a lab energy of 350 MeV corresponds roughly to a Fermi momentum 
$k_F=2.0\;{\rm fm}^{-1}$.  Therefore, calculations of the 
$^3P_2$-$^3F_2$ energy gap at densities above $k_F=2.0\;{\rm fm}^{-1}$ 
will inevitably involve extrapolating the potential models.  

In the same figure we also show the empirical $pp$ phase shifts 
obtained by Arndt et al.\ in a recent phase shift analysis \cite{arndt97}.  
Some differences between this phase shifts analysis and 
the phase shifts calculated with the potentials 
could be present in the figure, even below 350 MeV, 
because the potentials are not fitted to the analysis of Arndt et al..
The modern potentials fits the Nijmegen database \cite{nijpwa}, 
the older ones fit different analyses made in the 70's and 80's.  
Nevertheless, the four modern potentials considered 
here fit also Arndt analysis below 350 MeV with high accuracy,
while the old potentials (in particular the $V_{14}$)
overshoot the empirical values already at lower scattering energies,
due to the fact that they have a higher $\chi^2/{\rm datum}$ than 
the new models.

In any case, above $E_{\rm lab}=350\;{\rm MeV}$ 
(corresponding to $k_F\approx 2.0\;{\rm fm}^{-1}$)
sizeable differences show up in the predictions of all potentials.  
The Nijm-II potential fits the phase shifts 
up to about 600 MeV rather well, but after that it severely 
overestimates them.  
This in turn means that the high-momentum components of the 
$^3P_2$ interaction will be too attractive.  
Nijm-I does fairly well up to about 500 MeV, from 500 to 700 MeV it 
underpredicts the phase shifts, while at energies above 700 MeV the 
results are too high.  
The CD-Bonn potential gives a similar behavior, but falls 
faster towards zero at high energies than Nijm-I and II.  
Argonne $V_{18}$ gives $^3P_2$ phase shifts below the empirical ones 
over the whole range $E_{\rm lab}=400$--$1000\;{\rm MeV}$. 
The old potentials display similar variations, being generally too repulsive
with Paris the most repulsive of all potentials, followed by Bonn B and
Argonne $V_{14}$.
In this paper we will further on focus on the new, phase-shift 
equivalent potentials, since they are better fitted to modern scattering
data.      
In summary, all potentials give phase shifts which are too attractive 
above $E_{\rm lab}\approx 700$--$1000\;{\rm MeV}$, 
and all except Nijm-II are too repulsive between $\approx 350\;{\rm MeV}$ 
and $\approx 700$--$1000\;{\rm MeV}$.    

\section{Results}
\label{sec:results} 

Before presenting results for the energy gap, we  
point out some features of the gap equations which make the 
trend of the results understandable.  
In order to make the connection to the NN interaction as transparent 
as possible, we start by discussing the case where the 
single-particle energies are given by their values in free space, 
$e(k)=k^2/2m$. 
 
\begin{figure}
\setlength{\unitlength}{0.1bp}
\special{!
/gnudict 40 dict def
gnudict begin
/Color false def
/Solid false def
/gnulinewidth 5.000 def
/vshift -33 def
/dl {10 mul} def
/hpt 31.5 def
/vpt 31.5 def
/M {moveto} bind def
/L {lineto} bind def
/R {rmoveto} bind def
/V {rlineto} bind def
/vpt2 vpt 2 mul def
/hpt2 hpt 2 mul def
/Lshow { currentpoint stroke M
  0 vshift R show } def
/Rshow { currentpoint stroke M
  dup stringwidth pop neg vshift R show } def
/Cshow { currentpoint stroke M
  dup stringwidth pop -2 div vshift R show } def
/DL { Color {setrgbcolor Solid {pop []} if 0 setdash }
 {pop pop pop Solid {pop []} if 0 setdash} ifelse } def
/BL { stroke gnulinewidth 2 mul setlinewidth } def
/AL { stroke gnulinewidth 2 div setlinewidth } def
/PL { stroke gnulinewidth setlinewidth } def
/LTb { BL [] 0 0 0 DL } def
/LTa { AL [1 dl 2 dl] 0 setdash 0 0 0 setrgbcolor } def
/LT0 { PL [] 0 1 0 DL } def
/LT1 { PL [4 dl 2 dl] 0 0 1 DL } def
/LT2 { PL [2 dl 3 dl] 1 0 0 DL } def
/LT3 { PL [1 dl 1.5 dl] 1 0 1 DL } def
/LT4 { PL [5 dl 2 dl 1 dl 2 dl] 0 1 1 DL } def
/LT5 { PL [4 dl 3 dl 1 dl 3 dl] 1 1 0 DL } def
/LT6 { PL [2 dl 2 dl 2 dl 4 dl] 0 0 0 DL } def
/LT7 { PL [2 dl 2 dl 2 dl 2 dl 2 dl 4 dl] 1 0.3 0 DL } def
/LT8 { PL [2 dl 2 dl 2 dl 2 dl 2 dl 2 dl 2 dl 4 dl] 0.5 0.5 0.5 DL } def
/P { stroke [] 0 setdash
  currentlinewidth 2 div sub M
  0 currentlinewidth V stroke } def
/D { stroke [] 0 setdash 2 copy vpt add M
  hpt neg vpt neg V hpt vpt neg V
  hpt vpt V hpt neg vpt V closepath stroke
  P } def
/A { stroke [] 0 setdash vpt sub M 0 vpt2 V
  currentpoint stroke M
  hpt neg vpt neg R hpt2 0 V stroke
  } def
/B { stroke [] 0 setdash 2 copy exch hpt sub exch vpt add M
  0 vpt2 neg V hpt2 0 V 0 vpt2 V
  hpt2 neg 0 V closepath stroke
  P } def
/C { stroke [] 0 setdash exch hpt sub exch vpt add M
  hpt2 vpt2 neg V currentpoint stroke M
  hpt2 neg 0 R hpt2 vpt2 V stroke } def
/T { stroke [] 0 setdash 2 copy vpt 1.12 mul add M
  hpt neg vpt -1.62 mul V
  hpt 2 mul 0 V
  hpt neg vpt 1.62 mul V closepath stroke
  P  } def
/S { 2 copy A C} def
end
}
\begin{picture}(3600,2160)(0,0)
\special{"
gnudict begin
gsave
50 50 translate
0.100 0.100 scale
0 setgray
/Helvetica findfont 100 scalefont setfont
newpath
-500.000000 -500.000000 translate
LTa
600 251 M
2817 0 V
600 251 M
0 1858 V
LTb
600 251 M
63 0 V
2754 0 R
-63 0 V
600 499 M
63 0 V
2754 0 R
-63 0 V
600 746 M
63 0 V
2754 0 R
-63 0 V
600 994 M
63 0 V
2754 0 R
-63 0 V
600 1242 M
63 0 V
2754 0 R
-63 0 V
600 1490 M
63 0 V
2754 0 R
-63 0 V
600 1737 M
63 0 V
2754 0 R
-63 0 V
600 1985 M
63 0 V
2754 0 R
-63 0 V
600 251 M
0 63 V
0 1795 R
0 -63 V
882 251 M
0 63 V
0 1795 R
0 -63 V
1163 251 M
0 63 V
0 1795 R
0 -63 V
1445 251 M
0 63 V
0 1795 R
0 -63 V
1727 251 M
0 63 V
0 1795 R
0 -63 V
2009 251 M
0 63 V
0 1795 R
0 -63 V
2290 251 M
0 63 V
0 1795 R
0 -63 V
2572 251 M
0 63 V
0 1795 R
0 -63 V
2854 251 M
0 63 V
0 1795 R
0 -63 V
3135 251 M
0 63 V
0 1795 R
0 -63 V
3417 251 M
0 63 V
0 1795 R
0 -63 V
600 251 M
2817 0 V
0 1858 V
-2817 0 V
600 251 L
LT0
3114 1946 M
180 0 V
600 251 M
1 0 V
1 0 V
2 0 V
3 0 V
3 0 V
4 0 V
4 0 V
5 0 V
5 0 V
6 0 V
7 0 V
7 0 V
7 0 V
9 0 V
8 0 V
9 0 V
10 0 V
10 0 V
10 0 V
11 0 V
12 0 V
11 0 V
13 0 V
12 0 V
13 0 V
13 0 V
14 0 V
13 0 V
14 0 V
14 0 V
15 0 V
15 0 V
14 0 V
15 1 V
15 0 V
16 0 V
15 0 V
15 0 V
16 0 V
15 0 V
15 0 V
16 0 V
15 1 V
15 0 V
15 0 V
15 0 V
15 1 V
15 0 V
14 0 V
14 1 V
14 0 V
14 0 V
14 1 V
13 0 V
13 1 V
12 0 V
12 1 V
12 1 V
11 0 V
11 1 V
11 1 V
10 1 V
9 1 V
10 1 V
8 1 V
8 2 V
8 1 V
7 1 V
7 2 V
6 2 V
5 2 V
5 1 V
4 2 V
4 2 V
3 1 V
3 2 V
1 1 V
2 1 V
1 0 V
0 1 V
2 1 V
2 2 V
4 3 V
4 4 V
4 5 V
4 6 V
3 6 V
3 6 V
2 3 V
0 2 V
2 5 V
2 9 V
4 15 V
4 26 V
4 46 V
4 85 V
4 182 V
2 398 V
2 398 V
0 2 V
2 -394 V
3 -397 V
3 -182 V
4 -86 V
4 -45 V
4 -26 V
4 -15 V
2 -9 V
2 -5 V
1 -2 V
1 -4 V
3 -5 V
3 -6 V
4 -6 V
4 -5 V
4 -4 V
4 -3 V
3 -2 V
1 -1 V
3 -2 V
10 -6 V
17 -6 V
26 -6 V
33 -4 V
41 -4 V
49 -2 V
57 -2 V
64 -2 V
71 -1 V
79 -1 V
86 -1 V
93 0 V
100 -1 V
107 -1 V
113 0 V
120 0 V
126 -1 V
132 0 V
138 0 V
144 -1 V
149 0 V
155 0 V
3 0 V
LT1
3114 1846 M
180 0 V
600 251 M
2 0 V
2 0 V
3 0 V
5 0 V
5 0 V
6 0 V
8 0 V
8 0 V
9 0 V
11 0 V
11 0 V
12 0 V
13 0 V
14 0 V
15 0 V
15 0 V
17 0 V
17 0 V
18 0 V
19 0 V
19 0 V
20 0 V
21 0 V
22 0 V
22 0 V
22 0 V
23 0 V
24 0 V
24 1 V
24 0 V
25 0 V
25 0 V
25 0 V
26 0 V
26 1 V
26 0 V
26 0 V
27 0 V
26 1 V
26 0 V
27 1 V
26 0 V
26 1 V
26 0 V
26 1 V
26 0 V
25 1 V
25 1 V
25 1 V
24 1 V
24 1 V
24 1 V
23 1 V
23 2 V
22 1 V
21 2 V
21 1 V
20 2 V
19 2 V
19 3 V
18 3 V
17 2 V
17 4 V
15 3 V
15 5 V
14 4 V
13 5 V
12 6 V
12 6 V
10 7 V
9 8 V
8 8 V
8 8 V
6 9 V
6 9 V
4 8 V
3 7 V
3 5 V
1 3 V
1 2 V
1 4 V
3 7 V
3 11 V
4 15 V
4 18 V
4 21 V
4 23 V
2 20 V
2 13 V
1 7 V
1 16 V
3 31 V
3 49 V
4 82 V
5 128 V
4 192 V
3 237 V
3 175 V
1 50 V
1 2 V
1 -46 V
3 -171 V
3 -236 V
4 -192 V
5 -129 V
4 -82 V
3 -51 V
3 -31 V
1 -15 V
1 -7 V
2 -13 V
2 -20 V
4 -23 V
4 -21 V
4 -18 V
4 -15 V
3 -11 V
3 -7 V
1 -4 V
3 -6 V
9 -19 V
17 -23 V
24 -21 V
32 -17 V
39 -12 V
46 -10 V
53 -7 V
61 -5 V
67 -5 V
75 -3 V
81 -3 V
88 -2 V
95 -2 V
101 -2 V
107 -1 V
113 -2 V
120 -1 V
125 0 V
97 -1 V
LT2
3114 1746 M
180 0 V
600 251 M
2 0 V
4 0 V
4 0 V
6 0 V
8 0 V
9 0 V
10 0 V
12 0 V
14 0 V
14 0 V
16 0 V
17 0 V
19 0 V
19 0 V
21 0 V
22 0 V
24 0 V
24 0 V
26 0 V
26 0 V
28 0 V
28 0 V
30 0 V
30 0 V
31 0 V
32 0 V
33 0 V
33 0 V
34 0 V
35 0 V
35 0 V
35 1 V
36 0 V
37 0 V
36 0 V
37 0 V
37 0 V
38 0 V
37 1 V
37 0 V
38 0 V
37 0 V
37 0 V
37 1 V
37 0 V
36 0 V
36 1 V
36 0 V
35 1 V
34 0 V
34 1 V
34 0 V
32 1 V
32 1 V
31 0 V
31 1 V
29 1 V
29 1 V
27 1 V
27 2 V
25 1 V
25 1 V
23 2 V
22 2 V
21 2 V
20 3 V
18 3 V
18 3 V
15 4 V
15 4 V
13 5 V
12 5 V
11 6 V
9 7 V
7 6 V
6 7 V
5 5 V
3 5 V
2 3 V
1 1 V
1 2 V
3 5 V
4 7 V
4 9 V
4 12 V
4 14 V
3 14 V
3 13 V
1 9 V
1 4 V
2 11 V
2 20 V
4 33 V
4 56 V
4 96 V
4 163 V
3 267 V
3 298 V
2 116 V
0 1 V
2 -114 V
2 -297 V
4 -268 V
4 -164 V
4 -96 V
4 -57 V
4 -34 V
2 -21 V
2 -10 V
0 -4 V
2 -9 V
3 -13 V
3 -14 V
4 -14 V
4 -11 V
4 -10 V
4 -6 V
2 -5 V
2 -3 V
2 -3 V
9 -11 V
16 -14 V
23 -13 V
29 -11 V
37 -8 V
44 -6 V
50 -4 V
57 -4 V
64 -3 V
70 -2 V
77 -1 V
83 -2 V
89 -1 V
95 -1 V
44 0 V
stroke
grestore
end
showpage
}
\put(3054,1746){\makebox(0,0)[r]{$k_F=3.5\;{\rm fm}^{-1}$}}
\put(3054,1846){\makebox(0,0)[r]{$k_F=2.5\;{\rm fm}^{-1}$}}
\put(3054,1946){\makebox(0,0)[r]{$k_F=1.5\;{\rm fm}^{-1}$}}
\put(2008,51){\makebox(0,0){$k$ (${\rm fm}^{-1}$) }}
\put(100,1180){%
\special{ps: gsave currentpoint currentpoint translate
270 rotate neg exch neg exch translate}%
\makebox(0,0)[b]{\shortstack{Normalized integrand (dimensionless)}}%
\special{ps: currentpoint grestore moveto}%
}
\put(3417,151){\makebox(0,0){5}}
\put(3135,151){\makebox(0,0){4.5}}
\put(2854,151){\makebox(0,0){4}}
\put(2572,151){\makebox(0,0){3.5}}
\put(2290,151){\makebox(0,0){3}}
\put(2009,151){\makebox(0,0){2.5}}
\put(1727,151){\makebox(0,0){2}}
\put(1445,151){\makebox(0,0){1.5}}
\put(1163,151){\makebox(0,0){1}}
\put(882,151){\makebox(0,0){0.5}}
\put(600,151){\makebox(0,0){0}}
\put(540,1985){\makebox(0,0)[r]{1.4}}
\put(540,1737){\makebox(0,0)[r]{1.2}}
\put(540,1490){\makebox(0,0)[r]{1}}
\put(540,1242){\makebox(0,0)[r]{0.8}}
\put(540,994){\makebox(0,0)[r]{0.6}}
\put(540,746){\makebox(0,0)[r]{0.4}}
\put(540,499){\makebox(0,0)[r]{0.2}}
\put(540,251){\makebox(0,0)[r]{0}}
\end{picture}
\caption{$^3P_2$ part of the integrand in the gap equations for 
         various densities and with the Nijm-I potential.}
\label{fig:int}
\end{figure}
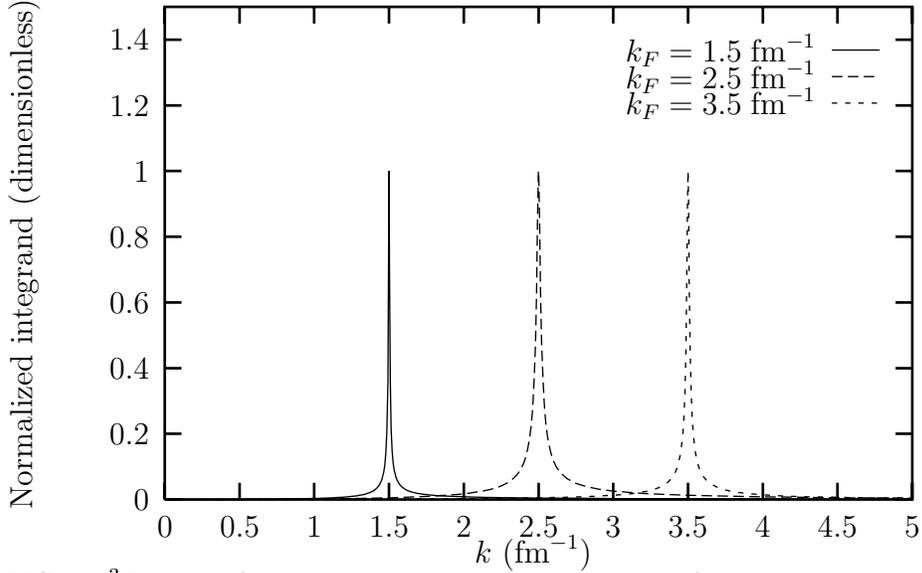

In Fig.~\ref{fig:int} we show, 
for the Nijm-I potential and various values of $k_F$, 
the function $k^2\Delta_1(k)/E(k)$ 
involved in the $^3P_2$ component of the gap equations, 
normalized to unity at $k=k_F$.
The behavior of this function was found to be the same for all potentials.  
Notice that this function is 
very strongly peaked around $k=k_F$, implying that 
the diagonal matrix element of the potential at $k=k_F$ 
gives the most important contribution to 
$\Delta_1(k_F)$ and $\Delta_3(k_F)$.  
Also, this figure makes it clear why some care in choosing 
momentum mesh points for the numerical integrations is needed.  
The function $k^2\Delta_3(k)/E(k)$ shows a similar, strongly 
peaked behavior, and thus the gap is largely determined by 
the matrix elements $V_{11}(k_F,k_F)$, $V_{13}(k_F,k_F)$ and 
$V_{33}(k_F,k_F)$.  

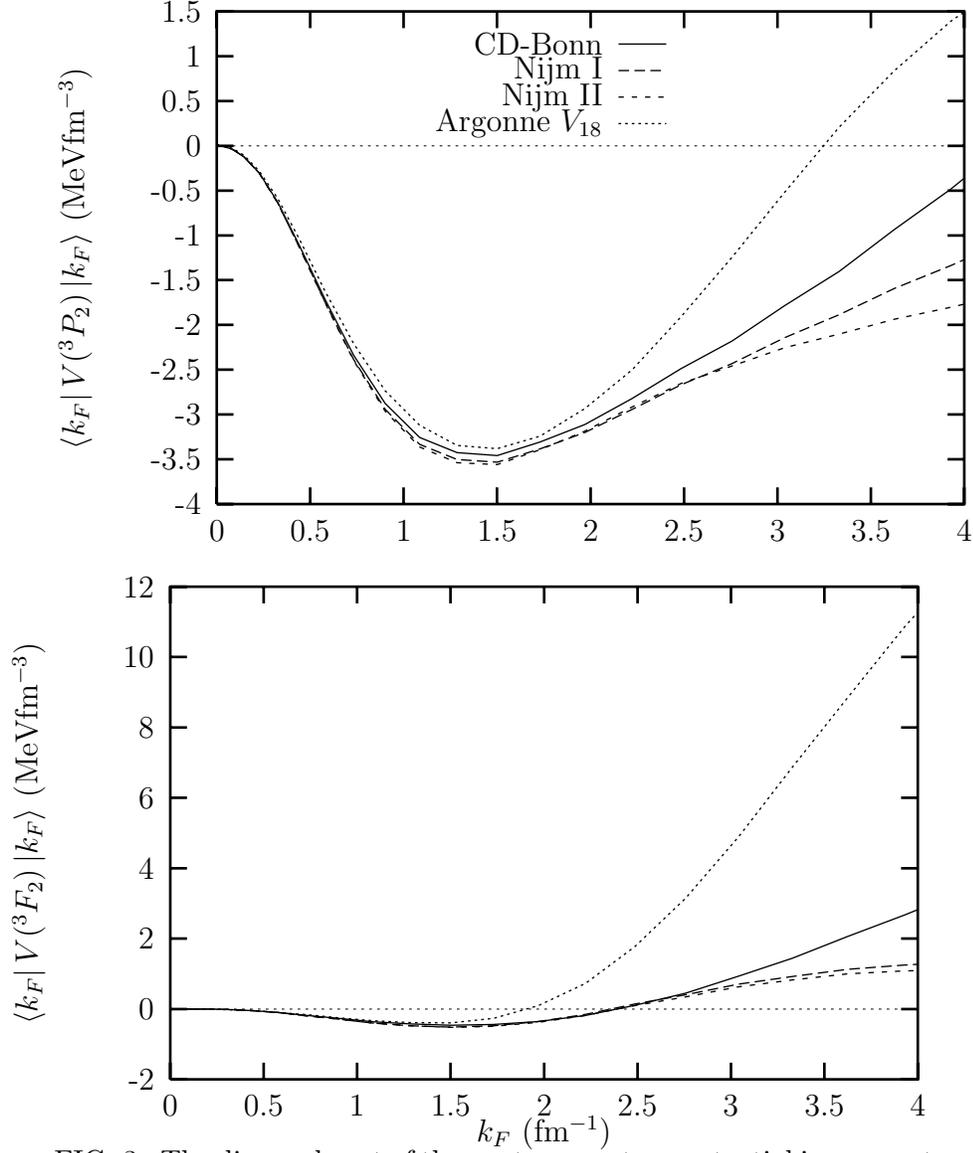
\begin{figure}
\setlength{\unitlength}{0.1bp}
\special{!
/gnudict 40 dict def
gnudict begin
/Color false def
/Solid false def
/gnulinewidth 5.000 def
/vshift -33 def
/dl {10 mul} def
/hpt 31.5 def
/vpt 31.5 def
/M {moveto} bind def
/L {lineto} bind def
/R {rmoveto} bind def
/V {rlineto} bind def
/vpt2 vpt 2 mul def
/hpt2 hpt 2 mul def
/Lshow { currentpoint stroke M
  0 vshift R show } def
/Rshow { currentpoint stroke M
  dup stringwidth pop neg vshift R show } def
/Cshow { currentpoint stroke M
  dup stringwidth pop -2 div vshift R show } def
/DL { Color {setrgbcolor Solid {pop []} if 0 setdash }
 {pop pop pop Solid {pop []} if 0 setdash} ifelse } def
/BL { stroke gnulinewidth 2 mul setlinewidth } def
/AL { stroke gnulinewidth 2 div setlinewidth } def
/PL { stroke gnulinewidth setlinewidth } def
/LTb { BL [] 0 0 0 DL } def
/LTa { AL [1 dl 2 dl] 0 setdash 0 0 0 setrgbcolor } def
/LT0 { PL [] 0 1 0 DL } def
/LT1 { PL [4 dl 2 dl] 0 0 1 DL } def
/LT2 { PL [2 dl 3 dl] 1 0 0 DL } def
/LT3 { PL [1 dl 1.5 dl] 1 0 1 DL } def
/LT4 { PL [5 dl 2 dl 1 dl 2 dl] 0 1 1 DL } def
/LT5 { PL [4 dl 3 dl 1 dl 3 dl] 1 1 0 DL } def
/LT6 { PL [2 dl 2 dl 2 dl 4 dl] 0 0 0 DL } def
/LT7 { PL [2 dl 2 dl 2 dl 2 dl 2 dl 4 dl] 1 0.3 0 DL } def
/LT8 { PL [2 dl 2 dl 2 dl 2 dl 2 dl 2 dl 2 dl 4 dl] 0.5 0.5 0.5 DL } def
/P { stroke [] 0 setdash
  currentlinewidth 2 div sub M
  0 currentlinewidth V stroke } def
/D { stroke [] 0 setdash 2 copy vpt add M
  hpt neg vpt neg V hpt vpt neg V
  hpt vpt V hpt neg vpt V closepath stroke
  P } def
/A { stroke [] 0 setdash vpt sub M 0 vpt2 V
  currentpoint stroke M
  hpt neg vpt neg R hpt2 0 V stroke
  } def
/B { stroke [] 0 setdash 2 copy exch hpt sub exch vpt add M
  0 vpt2 neg V hpt2 0 V 0 vpt2 V
  hpt2 neg 0 V closepath stroke
  P } def
/C { stroke [] 0 setdash exch hpt sub exch vpt add M
  hpt2 vpt2 neg V currentpoint stroke M
  hpt2 neg 0 R hpt2 vpt2 V stroke } def
/T { stroke [] 0 setdash 2 copy vpt 1.12 mul add M
  hpt neg vpt -1.62 mul V
  hpt 2 mul 0 V
  hpt neg vpt 1.62 mul V closepath stroke
  P  } def
/S { 2 copy A C} def
end
}
\begin{picture}(3600,2160)(0,0)
\special{"
gnudict begin
gsave
50 50 translate
0.100 0.100 scale
0 setgray
/Helvetica findfont 100 scalefont setfont
newpath
-500.000000 -500.000000 translate
LTa
600 1602 M
2817 0 V
600 251 M
0 1858 V
LTb
600 251 M
63 0 V
2754 0 R
-63 0 V
600 420 M
63 0 V
2754 0 R
-63 0 V
600 589 M
63 0 V
2754 0 R
-63 0 V
600 758 M
63 0 V
2754 0 R
-63 0 V
600 927 M
63 0 V
2754 0 R
-63 0 V
600 1096 M
63 0 V
2754 0 R
-63 0 V
600 1264 M
63 0 V
2754 0 R
-63 0 V
600 1433 M
63 0 V
2754 0 R
-63 0 V
600 1602 M
63 0 V
2754 0 R
-63 0 V
600 1771 M
63 0 V
2754 0 R
-63 0 V
600 1940 M
63 0 V
2754 0 R
-63 0 V
600 2109 M
63 0 V
2754 0 R
-63 0 V
600 251 M
0 63 V
0 1795 R
0 -63 V
952 251 M
0 63 V
0 1795 R
0 -63 V
1304 251 M
0 63 V
0 1795 R
0 -63 V
1656 251 M
0 63 V
0 1795 R
0 -63 V
2009 251 M
0 63 V
0 1795 R
0 -63 V
2361 251 M
0 63 V
0 1795 R
0 -63 V
2713 251 M
0 63 V
0 1795 R
0 -63 V
3065 251 M
0 63 V
0 1795 R
0 -63 V
3417 251 M
0 63 V
0 1795 R
0 -63 V
600 251 M
2817 0 V
0 1858 V
-2817 0 V
600 251 L
LT0
2114 1986 M
180 0 V
600 1602 M
1 0 V
1 0 V
1 0 V
1 0 V
1 0 V
1 0 V
1 0 V
4 0 V
17 -3 V
31 -10 V
43 -28 V
58 -61 V
70 -111 V
83 -166 V
95 -200 V
1116 812 L
1235 630 L
1365 502 L
141 -57 V
151 -11 V
161 51 V
169 66 V
178 98 V
185 113 V
192 104 V
199 134 V
204 128 V
208 158 V
212 152 V
52 42 V
LT1
2114 1886 M
180 0 V
600 1602 M
1 0 V
1 0 V
1 0 V
1 0 V
1 0 V
1 0 V
1 0 V
4 0 V
17 -3 V
31 -10 V
43 -28 V
58 -62 V
70 -113 V
83 -170 V
95 -203 V
1116 796 L
1235 608 L
1365 476 L
141 -57 V
151 -10 V
161 50 V
169 62 V
178 87 V
185 93 V
192 80 V
199 98 V
204 86 V
208 99 V
212 86 V
52 23 V
LT2
2114 1786 M
180 0 V
600 1602 M
1 0 V
1 0 V
1 0 V
1 0 V
1 0 V
1 0 V
1 0 V
4 0 V
17 -3 V
31 -10 V
43 -28 V
58 -62 V
70 -113 V
83 -170 V
95 -205 V
1116 792 L
1235 601 L
1365 466 L
141 -59 V
151 -7 V
161 56 V
169 70 V
178 91 V
185 89 V
192 66 V
199 69 V
204 51 V
208 56 V
212 45 V
52 12 V
LT3
2114 1686 M
180 0 V
600 1606 M
1 0 V
1 0 V
1 0 V
1 0 V
0 -1 V
1 0 V
1 0 V
1 0 V
4 0 V
17 -3 V
31 -9 V
43 -26 V
58 -60 V
70 -104 V
83 -156 V
95 -193 V
1116 856 L
1235 678 L
1365 548 L
141 -76 V
151 -12 V
161 47 V
169 104 V
178 146 V
185 201 V
192 224 V
199 246 V
204 244 V
208 213 V
212 186 V
52 36 V
stroke
grestore
end
showpage
}
\put(2054,1686){\makebox(0,0)[r]{Argonne $V_{18}$}}
\put(2054,1786){\makebox(0,0)[r]{Nijm II}}
\put(2054,1886){\makebox(0,0)[r]{Nijm I}}
\put(2054,1986){\makebox(0,0)[r]{CD-Bonn}}
\put(100,1180){%
\special{ps: gsave currentpoint currentpoint translate
270 rotate neg exch neg exch translate}%
\makebox(0,0)[b]{\shortstack{$\left\langle k_F \right| V(^3P_2)\left| k_F\right\rangle$ (MeVfm$^{-3}$)}}%
\special{ps: currentpoint grestore moveto}%
}
\put(3417,151){\makebox(0,0){4}}
\put(3065,151){\makebox(0,0){3.5}}
\put(2713,151){\makebox(0,0){3}}
\put(2361,151){\makebox(0,0){2.5}}
\put(2009,151){\makebox(0,0){2}}
\put(1656,151){\makebox(0,0){1.5}}
\put(1304,151){\makebox(0,0){1}}
\put(952,151){\makebox(0,0){0.5}}
\put(600,151){\makebox(0,0){0}}
\put(540,2109){\makebox(0,0)[r]{1.5}}
\put(540,1940){\makebox(0,0)[r]{1}}
\put(540,1771){\makebox(0,0)[r]{0.5}}
\put(540,1602){\makebox(0,0)[r]{0}}
\put(540,1433){\makebox(0,0)[r]{-0.5}}
\put(540,1264){\makebox(0,0)[r]{-1}}
\put(540,1096){\makebox(0,0)[r]{-1.5}}
\put(540,927){\makebox(0,0)[r]{-2}}
\put(540,758){\makebox(0,0)[r]{-2.5}}
\put(540,589){\makebox(0,0)[r]{-3}}
\put(540,420){\makebox(0,0)[r]{-3.5}}
\put(540,251){\makebox(0,0)[r]{-4}}
\end{picture}
\begin{picture}(3600,2160)(0,0)
\special{"
gnudict begin
gsave
50 50 translate
0.100 0.100 scale
0 setgray
/Helvetica findfont 100 scalefont setfont
newpath
-500.000000 -500.000000 translate
LTa
600 516 M
2817 0 V
600 251 M
0 1858 V
LTb
600 251 M
63 0 V
2754 0 R
-63 0 V
600 516 M
63 0 V
2754 0 R
-63 0 V
600 782 M
63 0 V
2754 0 R
-63 0 V
600 1047 M
63 0 V
2754 0 R
-63 0 V
600 1313 M
63 0 V
2754 0 R
-63 0 V
600 1578 M
63 0 V
2754 0 R
-63 0 V
600 1844 M
63 0 V
2754 0 R
-63 0 V
600 2109 M
63 0 V
2754 0 R
-63 0 V
600 251 M
0 63 V
0 1795 R
0 -63 V
952 251 M
0 63 V
0 1795 R
0 -63 V
1304 251 M
0 63 V
0 1795 R
0 -63 V
1656 251 M
0 63 V
0 1795 R
0 -63 V
2009 251 M
0 63 V
0 1795 R
0 -63 V
2361 251 M
0 63 V
0 1795 R
0 -63 V
2713 251 M
0 63 V
0 1795 R
0 -63 V
3065 251 M
0 63 V
0 1795 R
0 -63 V
3417 251 M
0 63 V
0 1795 R
0 -63 V
600 251 M
2817 0 V
0 1858 V
-2817 0 V
600 251 L
LT0
600 516 M
1 0 V
1 0 V
1 0 V
1 0 V
1 0 V
1 0 V
1 0 V
4 0 V
17 0 V
31 0 V
43 0 V
58 0 V
70 -2 V
83 -4 V
95 -7 V
108 -11 V
119 -11 V
130 -13 V
141 -8 V
151 -5 V
161 3 V
169 11 V
178 23 V
185 38 V
192 46 V
199 64 V
204 68 V
208 82 V
212 80 V
52 21 V
LT1
600 516 M
1 0 V
1 0 V
1 0 V
1 0 V
1 0 V
1 0 V
1 0 V
4 0 V
17 0 V
31 0 V
43 0 V
58 0 V
70 -2 V
83 -4 V
95 -7 V
108 -13 V
119 -12 V
130 -15 V
141 -10 V
151 -5 V
161 5 V
169 16 V
178 27 V
185 39 V
192 35 V
199 40 V
204 29 V
208 27 V
212 16 V
52 3 V
LT2
600 516 M
1 0 V
1 0 V
1 0 V
1 0 V
1 0 V
1 0 V
1 0 V
4 0 V
17 0 V
31 0 V
43 0 V
58 0 V
70 -2 V
83 -4 V
95 -7 V
108 -13 V
119 -12 V
130 -15 V
141 -10 V
151 -6 V
161 4 V
169 15 V
178 26 V
185 37 V
192 33 V
199 38 V
204 26 V
208 23 V
212 11 V
52 1 V
LT3
600 515 M
1 0 V
1 0 V
1 0 V
1 0 V
1 0 V
1 0 V
1 0 V
0 1 V
4 0 V
17 0 V
31 0 V
43 -1 V
58 0 V
70 -2 V
83 -3 V
95 -7 V
108 -8 V
119 -12 V
130 -11 V
141 -7 V
151 -1 V
161 17 V
169 48 V
178 84 V
185 138 V
192 183 V
199 231 V
204 264 V
208 262 V
212 264 V
52 58 V
stroke
grestore
end
showpage
}
\put(2008,51){\makebox(0,0){$k_F$ (fm$^{-1}$)}}
\put(100,1180){%
\special{ps: gsave currentpoint currentpoint translate
270 rotate neg exch neg exch translate}%
\makebox(0,0)[b]{\shortstack{$\left\langle k_F \right| V(^3F_2)\left| k_F\right\rangle$ (MeVfm$^{-3}$)}}%
\special{ps: currentpoint grestore moveto}%
}
\put(3417,151){\makebox(0,0){4}}
\put(3065,151){\makebox(0,0){3.5}}
\put(2713,151){\makebox(0,0){3}}
\put(2361,151){\makebox(0,0){2.5}}
\put(2009,151){\makebox(0,0){2}}
\put(1656,151){\makebox(0,0){1.5}}
\put(1304,151){\makebox(0,0){1}}
\put(952,151){\makebox(0,0){0.5}}
\put(600,151){\makebox(0,0){0}}
\put(540,2109){\makebox(0,0)[r]{12}}
\put(540,1844){\makebox(0,0)[r]{10}}
\put(540,1578){\makebox(0,0)[r]{8}}
\put(540,1313){\makebox(0,0)[r]{6}}
\put(540,1047){\makebox(0,0)[r]{4}}
\put(540,782){\makebox(0,0)[r]{2}}
\put(540,516){\makebox(0,0)[r]{0}}
\put(540,251){\makebox(0,0)[r]{-2}}
\end{picture}
\caption{The diagonal part of the neutron-neutron potential in 
         momentum space [Eq.~(\ref{e:v})] 
         for the $^3P_2$ (top panel) and the $^3F_2$ (bottom panel) 
         partial waves obtained with the 
          CD-Bonn, Nijmegen I and II and Argonne $V_{18}$ potentials.}
\label{fig:pots}
\end{figure}

To exemplify this,
we have therefore plotted in Fig.~\ref{fig:pots} 
the matrix elements for 
$V_{11}(k_F,k_F)$ and $V_{33}(k_F,k_F)$ 
as functions of $k_F$ for the various modern 
potentials used in this work. 
Up to $k_F\approx 2.0\;{\rm fm}^{-1}$ the matrix elements are very similar,
but after this point they deviate from each other,  
in line with the phase shift predictions shown in Fig.~\ref{fig:phaseshift}: 
In the $^3P_2$ and $^3F_2$ waves, 
the $V_{18}$ potential is the most repulsive,
followed by the CD-Bonn and the Nijm-I and Nijm-II potentials in that order.
Similar conclusions can be reached for  
the coupled $^3P_2$-$^3F_2$ channel. 

\subsection{Pairing gaps}

Fig.~\ref{fig:gaps} contains a comprehensive collection of our results for
the pairing gaps with the different potentials. 
We start with the top part of the figure, which displays the results
calculated with free single-particle energies.  
Differences between the results are therefore solely due to differences 
in the $^3P_2$-$^3F_2$ matrix elements of the potentials.
The plot shows results obtained with the old as well as with the modern
potentials.
The results (with the notable exception of the Argonne 
$V_{14}$\footnote{
  In a previous paper \cite{bcll92}
  one of the authors (M.B.) has claimed much higher values for the
  gap with the Argonne $V_{14}$.  
  It has been checked that this was due both to a non accurate
  separable representation of the NN potential and to a bug in the
  computer program for this channel.},
which predicts also substantially different $^3P_2$ phase shifts
(see Fig.~\ref{fig:phaseshift}),
are in good agreement at densities below $k_F\approx 2.0\;{\rm fm}^{-1}$, 
but differ significantly at higher densities.  
This is in accordance with the fact that 
the diagonal matrix elements of the potentials are very similar 
below $k_F\approx 2.0\;{\rm fm}^{-1}$, corresponding  
to a laboratory energy for free NN scattering 
of $E_{\rm lab} \approx 350\;{\rm MeV}$.  
This indicates that
within this range the good fit of the potentials 
to scattering data below 350 MeV makes the ambiguities in the 
results for the energy gap quite small, since, to a first approximation, see 
the discussion below, the pairing gap can be derived in terms of the 
phase shifts only.

However, we wish to calculate the gap also at densities above 
$k_F=2.0\;{\rm fm}^{-1}$.  
Then we need the various potentials at higher energies, 
outside of the range where they are fitted to scattering data.  
Thus there is no guarantee that 
the results will be independent of the model chosen, and in fact 
the figure shows that there are considerable differences 
between their predictions at high densities,
following precisely the trend observed in the phase shift predictions:
The Argonne $V_{18}$ is the most repulsive of the modern potentials,
followed by the CD-Bonn and Nijmegen I and II.
Most remarkable are the results obtained with Nijm-II: 
we find that the predicted gap 
continues to rise unrealistically even at $k_F \approx 3.5\;{\rm fm}^{-1}$, 
where the purely nucleonic description of matter surely breaks down.
From Table \ref{tab:tab1}, which contains a compilation of gaps 
for the various potentias, one sees that the improved fit of the 
new potentials to scattering data leads to better agreement 
in their predictions for the gap.   
Thus, the fact that these potentials have been fit with high precision 
to the same set of scattering data eliminates some of the 
ambiguities, and allows one to compare interactions in a way not 
possible with earlier models.   

Since the potentials fail to reproduce the measured phase shifts 
beyond $E_{\rm lab}=350\;{\rm MeV}$, the predictions for the $^3P_2$-$^3F_2$ 
energy gap in neutron matter cannot be trusted above 
$k_F\approx 2.0\;{\rm fm}^{-1}$.  
Therefore, the behavior of the $^3P_2$-$^3F_2$ energy gap at high densities 
should be considered as unknown, and cannot be obtained until potential models 
which fit the phase shifts in the inelastic region 
above $E_{\rm lab}=350\;{\rm MeV}$ are constructed.  
These potential models need the flexibility to 
include both the flat structure in the phase shifts above 600 MeV, 
due to the ${\rm NN}\rightarrow{\rm N}\Delta$ channel, as well as the 
rapid decrease to zero at $E_{\rm lab}\approx 1100\;{\rm MeV}$.  
 
We proceed now to the middle part of Fig.~\ref{fig:gaps}, where
the results for the energy gap using BHF single-particle energies are shown.  
For details on the BHF calculations, see, e.g., Ref.~\cite{jjm76}.  
From this figure, two trends are apparent:  
First, the reduction of the in-medium nucleon mass leads to a sizeable 
reduction of the $^3P_2$-$^3F_2$ energy gap, as observed in 
earlier calculations \cite{amu85,bcll92,taka93,elga96}.  
Secondly, the new NN interactions give again similar results 
at low densities, while beyond $k_F\approx 2.0\;{\rm fm}^{-1}$ 
the gaps differ, as in the case with free single-particle energies.   

\begin{figure}
\setlength{\unitlength}{0.1bp}
\special{!
/gnudict 40 dict def
gnudict begin
/Color false def
/Solid false def
/gnulinewidth 5.000 def
/vshift -33 def
/dl {10 mul} def
/hpt 31.5 def
/vpt 31.5 def
/M {moveto} bind def
/L {lineto} bind def
/R {rmoveto} bind def
/V {rlineto} bind def
/vpt2 vpt 2 mul def
/hpt2 hpt 2 mul def
/Lshow { currentpoint stroke M
  0 vshift R show } def
/Rshow { currentpoint stroke M
  dup stringwidth pop neg vshift R show } def
/Cshow { currentpoint stroke M
  dup stringwidth pop -2 div vshift R show } def
/DL { Color {setrgbcolor Solid {pop []} if 0 setdash }
 {pop pop pop Solid {pop []} if 0 setdash} ifelse } def
/BL { stroke gnulinewidth 2 mul setlinewidth } def
/AL { stroke gnulinewidth 2 div setlinewidth } def
/PL { stroke gnulinewidth setlinewidth } def
/LTb { BL [] 0 0 0 DL } def
/LTa { AL [1 dl 2 dl] 0 setdash 0 0 0 setrgbcolor } def
/LT0 { PL [] 0 1 0 DL } def
/LT1 { PL [4 dl 2 dl] 0 0 1 DL } def
/LT2 { PL [2 dl 3 dl] 1 0 0 DL } def
/LT3 { PL [1 dl 1.5 dl] 1 0 1 DL } def
/LT4 { PL [5 dl 2 dl 1 dl 2 dl] 0 1 1 DL } def
/LT5 { PL [4 dl 3 dl 1 dl 3 dl] 1 1 0 DL } def
/LT6 { PL [2 dl 2 dl 2 dl 4 dl] 0 0 0 DL } def
/LT7 { PL [2 dl 2 dl 2 dl 2 dl 2 dl 4 dl] 1 0.3 0 DL } def
/LT8 { PL [2 dl 2 dl 2 dl 2 dl 2 dl 2 dl 2 dl 4 dl] 0.5 0.5 0.5 DL } def
/P { stroke [] 0 setdash
  currentlinewidth 2 div sub M
  0 currentlinewidth V stroke } def
/D { stroke [] 0 setdash 2 copy vpt add M
  hpt neg vpt neg V hpt vpt neg V
  hpt vpt V hpt neg vpt V closepath stroke
  P } def
/A { stroke [] 0 setdash vpt sub M 0 vpt2 V
  currentpoint stroke M
  hpt neg vpt neg R hpt2 0 V stroke
  } def
/B { stroke [] 0 setdash 2 copy exch hpt sub exch vpt add M
  0 vpt2 neg V hpt2 0 V 0 vpt2 V
  hpt2 neg 0 V closepath stroke
  P } def
/C { stroke [] 0 setdash exch hpt sub exch vpt add M
  hpt2 vpt2 neg V currentpoint stroke M
  hpt2 neg 0 R hpt2 vpt2 V stroke } def
/T { stroke [] 0 setdash 2 copy vpt 1.12 mul add M
  hpt neg vpt -1.62 mul V
  hpt 2 mul 0 V
  hpt neg vpt 1.62 mul V closepath stroke
  P  } def
/S { 2 copy A C} def
end
}
\begin{picture}(3600,2160)(0,0)
\special{"
gnudict begin
gsave
50 50 translate
0.100 0.100 scale
0 setgray
/Helvetica findfont 100 scalefont setfont
newpath
-500.000000 -500.000000 translate
LTa
LTb
600 251 M
63 0 V
2754 0 R
-63 0 V
600 437 M
63 0 V
2754 0 R
-63 0 V
600 623 M
63 0 V
2754 0 R
-63 0 V
600 808 M
63 0 V
2754 0 R
-63 0 V
600 994 M
63 0 V
2754 0 R
-63 0 V
600 1180 M
63 0 V
2754 0 R
-63 0 V
600 1366 M
63 0 V
2754 0 R
-63 0 V
600 1552 M
63 0 V
2754 0 R
-63 0 V
600 1737 M
63 0 V
2754 0 R
-63 0 V
600 1923 M
63 0 V
2754 0 R
-63 0 V
600 2109 M
63 0 V
2754 0 R
-63 0 V
600 251 M
0 63 V
0 1795 R
0 -63 V
1033 251 M
0 63 V
0 1795 R
0 -63 V
1467 251 M
0 63 V
0 1795 R
0 -63 V
1900 251 M
0 63 V
0 1795 R
0 -63 V
2334 251 M
0 63 V
0 1795 R
0 -63 V
2767 251 M
0 63 V
0 1795 R
0 -63 V
3200 251 M
0 63 V
0 1795 R
0 -63 V
600 251 M
2817 0 V
0 1858 V
-2817 0 V
600 251 L
LT0
3114 1946 M
180 0 V
730 1483 M
2 -1 V
3 -1 V
4 -2 V
6 -2 V
7 -2 V
8 -3 V
9 -4 V
11 -4 V
12 -5 V
13 -5 V
15 -5 V
15 -6 V
17 -7 V
18 -7 V
20 -7 V
20 -8 V
22 -8 V
22 -9 V
24 -8 V
25 -10 V
26 -9 V
27 -10 V
28 -11 V
28 -10 V
30 -11 V
31 -11 V
32 -12 V
32 -11 V
33 -12 V
34 -11 V
35 -12 V
36 -12 V
36 -12 V
37 -12 V
37 -12 V
38 -12 V
39 -12 V
39 -12 V
39 -12 V
40 -12 V
41 -12 V
40 -12 V
41 -12 V
42 -11 V
41 -11 V
42 -11 V
42 -11 V
42 -11 V
42 -10 V
42 -10 V
41 -10 V
42 -10 V
42 -10 V
42 -9 V
41 -9 V
42 -9 V
41 -8 V
40 -8 V
41 -8 V
40 -7 V
39 -7 V
39 -6 V
39 -6 V
38 -6 V
37 -6 V
37 -5 V
36 -5 V
36 -4 V
35 -4 V
34 -4 V
33 -4 V
32 -2 V
32 -1 V
31 -2 V
30 -2 V
28 -2 V
28 -2 V
27 -2 V
26 -2 V
25 -2 V
24 -4 V
22 -5 V
22 -5 V
20 -5 V
20 -5 V
18 -5 V
17 -4 V
15 -5 V
15 -3 V
13 -4 V
12 -3 V
11 -2 V
9 -3 V
8 -2 V
7 -1 V
6 -2 V
4 -1 V
3 0 V
2 -1 V
LT1
3114 1846 M
180 0 V
730 1487 M
2 0 V
3 -1 V
4 -2 V
6 -2 V
7 -3 V
8 -3 V
9 -4 V
11 -4 V
12 -5 V
13 -6 V
15 -5 V
15 -7 V
17 -6 V
18 -8 V
20 -7 V
20 -8 V
22 -9 V
22 -9 V
24 -9 V
25 -10 V
26 -10 V
27 -11 V
28 -11 V
28 -11 V
30 -12 V
31 -12 V
32 -12 V
32 -13 V
33 -12 V
34 -13 V
35 -13 V
36 -14 V
36 -13 V
37 -14 V
37 -14 V
38 -14 V
39 -15 V
39 -14 V
39 -15 V
40 -15 V
41 -16 V
40 -15 V
41 -15 V
42 -15 V
41 -15 V
42 -15 V
42 -15 V
42 -16 V
42 -15 V
42 -15 V
41 -15 V
42 -15 V
42 -14 V
42 -15 V
41 -15 V
42 -15 V
41 -15 V
40 -14 V
41 -14 V
40 -15 V
39 -14 V
39 -14 V
39 -14 V
38 -14 V
37 -13 V
37 -14 V
36 -13 V
36 -13 V
35 -12 V
34 -13 V
33 -12 V
32 -10 V
32 -11 V
31 -10 V
30 -10 V
28 -10 V
28 -10 V
27 -10 V
26 -9 V
25 -10 V
24 -10 V
22 -12 V
22 -11 V
20 -11 V
20 -10 V
18 -10 V
17 -9 V
15 -9 V
15 -8 V
13 -7 V
12 -7 V
11 -6 V
9 -5 V
8 -4 V
7 -4 V
6 -3 V
4 -3 V
3 -1 V
2 -1 V
LT2
3114 1746 M
180 0 V
730 1467 M
2 -1 V
3 -1 V
4 -1 V
6 -3 V
7 -2 V
8 -3 V
9 -4 V
11 -4 V
12 -5 V
13 -5 V
15 -6 V
15 -6 V
17 -7 V
18 -7 V
20 -7 V
20 -8 V
22 -8 V
22 -9 V
24 -9 V
25 -9 V
26 -10 V
27 -10 V
28 -11 V
28 -10 V
30 -11 V
31 -11 V
32 -12 V
32 -11 V
33 -12 V
34 -12 V
35 -11 V
36 -12 V
36 -12 V
37 -12 V
37 -12 V
38 -12 V
39 -12 V
39 -12 V
39 -11 V
40 -12 V
41 -12 V
40 -11 V
41 -12 V
42 -11 V
41 -10 V
42 -11 V
42 -10 V
42 -9 V
42 -10 V
42 -9 V
41 -9 V
42 -8 V
42 -8 V
42 -8 V
41 -7 V
42 -7 V
41 -7 V
40 -6 V
41 -6 V
40 -5 V
39 -5 V
39 -5 V
39 -4 V
38 -4 V
37 -4 V
37 -4 V
36 -3 V
36 -4 V
35 -2 V
34 -3 V
33 -2 V
32 0 V
32 0 V
31 0 V
30 0 V
28 0 V
28 0 V
27 -1 V
26 0 V
25 -1 V
24 -2 V
22 -4 V
22 -4 V
20 -4 V
20 -4 V
18 -4 V
17 -3 V
15 -4 V
15 -2 V
13 -3 V
12 -2 V
11 -3 V
9 -1 V
8 -2 V
7 -1 V
6 -1 V
4 -1 V
3 0 V
2 -1 V
LT3
3114 1646 M
180 0 V
730 1461 M
2 -1 V
3 -1 V
4 -2 V
6 -2 V
7 -2 V
8 -4 V
9 -3 V
11 -4 V
12 -5 V
13 -5 V
15 -6 V
15 -6 V
17 -7 V
18 -7 V
20 -7 V
20 -8 V
22 -8 V
22 -9 V
24 -9 V
25 -10 V
26 -9 V
27 -10 V
28 -11 V
28 -11 V
30 -10 V
31 -12 V
32 -11 V
32 -11 V
33 -12 V
34 -11 V
35 -12 V
36 -12 V
36 -11 V
37 -12 V
37 -12 V
38 -11 V
39 -12 V
39 -12 V
39 -12 V
40 -11 V
41 -12 V
40 -11 V
41 -11 V
42 -10 V
41 -9 V
42 -9 V
42 -9 V
42 -8 V
42 -8 V
42 -7 V
41 -7 V
42 -6 V
42 -5 V
42 -5 V
41 -5 V
42 -4 V
41 -3 V
40 -3 V
41 -2 V
40 -1 V
39 -1 V
39 0 V
39 0 V
38 1 V
37 2 V
37 2 V
36 3 V
36 4 V
35 4 V
34 4 V
33 5 V
32 6 V
32 6 V
31 7 V
30 7 V
28 7 V
28 7 V
27 7 V
26 8 V
25 7 V
24 7 V
22 6 V
22 6 V
20 6 V
20 6 V
18 5 V
17 6 V
15 5 V
15 5 V
13 5 V
12 5 V
11 4 V
9 3 V
8 4 V
7 2 V
6 3 V
4 2 V
3 1 V
2 0 V
stroke
grestore
end
showpage
}
\put(3054,1646){\makebox(0,0)[r]{Nijm II}}
\put(3054,1746){\makebox(0,0)[r]{Nijm I}}
\put(3054,1846){\makebox(0,0)[r]{Argonne $V_{18}$}}
\put(3054,1946){\makebox(0,0)[r]{CD-Bonn}}
\put(2008,51){\makebox(0,0){$k_F$ (${\rm fm}^{-1}$) }}
\put(100,1180){%
\special{ps: gsave currentpoint currentpoint translate
270 rotate neg exch neg exch translate}%
\makebox(0,0)[b]{\shortstack{$m^*/m$ }}%
\special{ps: currentpoint grestore moveto}%
}
\put(3200,151){\makebox(0,0){2.4}}
\put(2767,151){\makebox(0,0){2.2}}
\put(2334,151){\makebox(0,0){2}}
\put(1900,151){\makebox(0,0){1.8}}
\put(1467,151){\makebox(0,0){1.6}}
\put(1033,151){\makebox(0,0){1.4}}
\put(600,151){\makebox(0,0){1.2}}
\put(540,2109){\makebox(0,0)[r]{1.2}}
\put(540,1923){\makebox(0,0)[r]{1.15}}
\put(540,1737){\makebox(0,0)[r]{1.1}}
\put(540,1552){\makebox(0,0)[r]{1.05}}
\put(540,1366){\makebox(0,0)[r]{1}}
\put(540,1180){\makebox(0,0)[r]{0.95}}
\put(540,994){\makebox(0,0)[r]{0.9}}
\put(540,808){\makebox(0,0)[r]{0.85}}
\put(540,623){\makebox(0,0)[r]{0.8}}
\put(540,437){\makebox(0,0)[r]{0.75}}
\put(540,251){\makebox(0,0)[r]{0.7}}
\end{picture}
\caption{Effective masses derived from various interactions in 
         the BHF approach.}
\label{fig:mstar}
\end{figure}
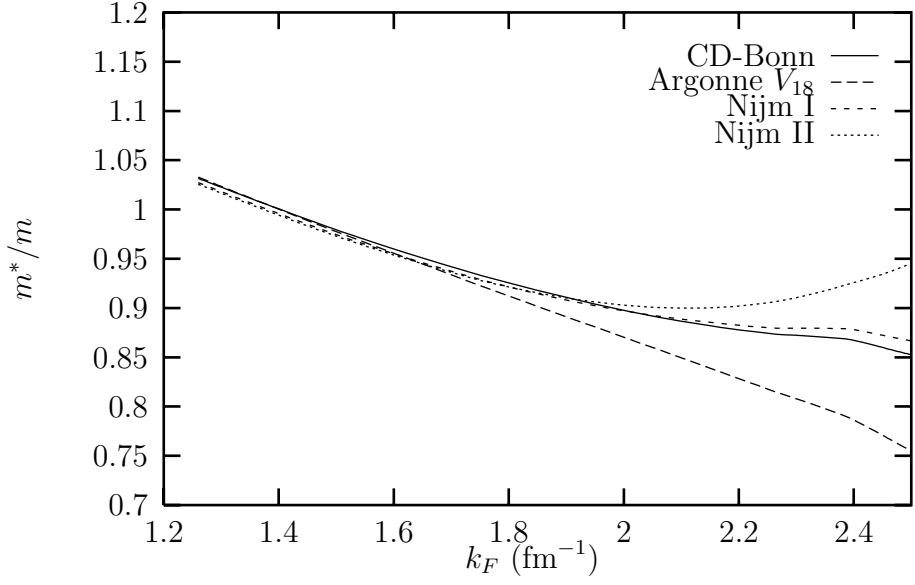

The single-particle energies at moderate densities obtained from the 
new potentials are rather similar, particularly in the 
important region near $k_F$.  
This is illustrated by a plot, Fig.~\ref{fig:mstar},
of the neutron effective mass,
\begin{equation}
 {m^*\over m} = \left( 1 + {m\over k_F} 
 \left.{dU\over dk}\right|_{k_F} \right)^{-1} \:,
\label{eq:effmass}
\end{equation}
as a function of density.
Up to $k_F \approx 2.0\;{\rm fm}^{-1}$ all results agree very closely,
but beyond that point the predictions diverge in the same manner as observed
for the phase shift predictions.
The differences of the BHF gaps at densities slightly above  
$k_F\approx 2.0\;{\rm fm}^{-1}$ are therefore mostly 
due to the differences in the $^3P_2$-$^3F_2$ waves of the potentials,
but at higher densities the differences between the gap are enhanced 
by differences in the single-particle potentials. The reader should bear in mind
that the single-particle energies contain
contributions from partial waves up to $l \leq 10$. The largest differences
arise however from contributions from the $^1S_0$ and  $^3P_2$-$^3F_2$
partial waves, see also the discussion in Ref.\ \cite{engv97}.
An extreme case is again the gap obtained with Nijm-II.  
It is caused by the very attractive $^3P_2$ matrix elements, 
amplified by the fact that the effective mass 
starts to increase at densities above $k_F\approx 2.5\;{\rm fm}^{-1}$ 
with this potential.  

Finally, in the lower panel of Fig.~\ref{fig:gaps}, we illustrate the
effect of different approximation schemes with an individual NN potential
(CD-Bonn), namely we compare the energy gaps obtained 
with the free single-particle spectrum, the BHF spectrum,
and an effective mass approximation,
\begin{equation}
  e(k) = U_0 + \frac{k^2}{2m^*} \:,
\label{eq:mstarapp}
\end{equation}
where $m^*$ is given in Eq.~(\ref{eq:effmass}).
In addition, also the gap in the uncoupled $^3P_2$ channel, 
i.e., neglecting the tensor coupling, is shown.

It becomes clear from Fig.~\ref{fig:gaps}  that the BHF spectrum forces a 
reduction of the gap by about a factor 2--3.
However, an effective mass aproximation should not be used when 
calculating the gap,  
because details of the single-particle spectrum around the Fermi 
momentum are important in order to obtain a correct value.   
The single-particle energies in the effective mass 
approximation are too steep near $k_F$.
We also emphasize that it is important to solve the coupled 
$^3P_2$-$^3F_2$ gap equations.  
By turning off the $^3P_2$-$^3F_2$ and $^3F_2$ channels, 
one obtains a $^3P_2$ gap that is considerably lower than the 
$^3P_2$-$^3F_2$ one.  
The reduction varies with the potential,
due to different strengths of the tensor force.
For more detailed discussions of the importance of the tensor force, 
the reader is referred to Refs.~\cite{amu85,taka93,elga96}.

\subsection{Hints from the $\bbox{^3P_2}$ phase shifts}

The first calculation of the $^3P_2$ gap in neutron matter was 
carried out by Hoffberg et al.~\cite{hoffberg70} in 1970.  
They used the weak-coupling expression for the energy gap to 
express it in terms of the $^3P_2$ phase shifts available at 
that time, and obtained 
a maximum gap of around 1 MeV at $k_F\approx 2.3\;{\rm fm}^{-1}$.    
Since all interactions considered in the present paper are 
fitted in the energy range 0--350 MeV, it would be interesting to 
use the recent phase shift analysis by Arndt et al.~\cite{arndt97} 
to get some hints on the behavior of the energy gap at higher densities.  
The phase shifts determine the interaction only on the energy shell, 
so to go from these ``experimental'' points to the energy gaps, 
we must make some rather strong assumptions. 
  
First of all, we switch off the interaction in the $^3F_2$ and 
$^3P_2$-$^3F_2$ channels and consider pure $^3P_2$ pairing.  
We are then left with only one gap equation to solve, and when we 
use the angle average approximation it is identical 
in form to the equation for $^1S_0$ pairing: 
\begin{equation}
  \Delta_1(k)=-\frac{1}{\pi}\int_{0}^{\infty}dk'k'^2V_{11}(k,k')
  \frac{\Delta_1(k')}{E(k')} \:.
\label{eq:sepeq1}
\end{equation}
In a recent paper \cite{eh98} two of the authors derived an 
expression for the $^1S_0$ gap in neutron and nuclear matter 
in terms of the phase shifts in this partial wave.  
This was possible because the interaction in this channel 
is to a good approximation rank-one separable at low energies 
due to the $^1S_0$ two-nucleon virtual state \cite{kkc96,carlson97}.  
No resonance or virtual state exists in the $^3P_2$ channel, but we will 
nevertheless approximate the interaction in this channel by 
a rank-one separable form, 
\begin{equation}
  V_{11}(k,k') = \lambda v(k)v(k') \:, 
\label{eq:sepeq2}
\end{equation}
where $\lambda$ is a constant.  
The interaction can then 
be expressed in terms of the phase shifts as \cite{eh98,jb76} 
\begin{equation}
  \lambda v^2(k) = -\frac{\sin\delta(k)}{k}e^{-\alpha(k)} \:, 
\label{eq:sepeq3}
\end{equation}
where $\alpha(k)$ is given by a principle value integral 
\begin{equation}
  \alpha(k)=\frac{1}{\pi}{\rm P}\int_{-\infty}^{+\infty}
  dk'\frac{\delta(k')}{k'-k} \:, 
\label{eq:sepeq4}
\end{equation}
and the phase shifts $\delta(k)$ are extended to negative momenta 
through $\delta(-k)=-\delta(k)$.  
This prescription works 
only for a purely attractive or purely repulsive interaction.  
The $^3P_2$ phase shifts change sign at $E_{\rm lab}\approx 1100\;{\rm MeV}$, 
and thus the interaction goes from attractive to repulsive at this energy.  
We therefore cut the integral in Eq.~(\ref{eq:sepeq4}) 
at $k\approx 3.6\;{\rm fm}^{-1}$, which corresponds to 
$E_{\rm lab}\approx 1100\;{\rm MeV}$.  
For a rank-one separable interaction, 
the solution of Eq.~(\ref{eq:sepeq1}) is given by 
$\Delta_F v(k)$, where $\Delta_F$ is the gap at the Fermi momentum 
found by solving 
\begin{equation}
 \frac{1}{\pi}\int_{0}^{\infty}dk'k'^2\frac{\lambda v^2(k')}{E(k')} = -1 \:.
\label{eq:sepeq5}
\end{equation}

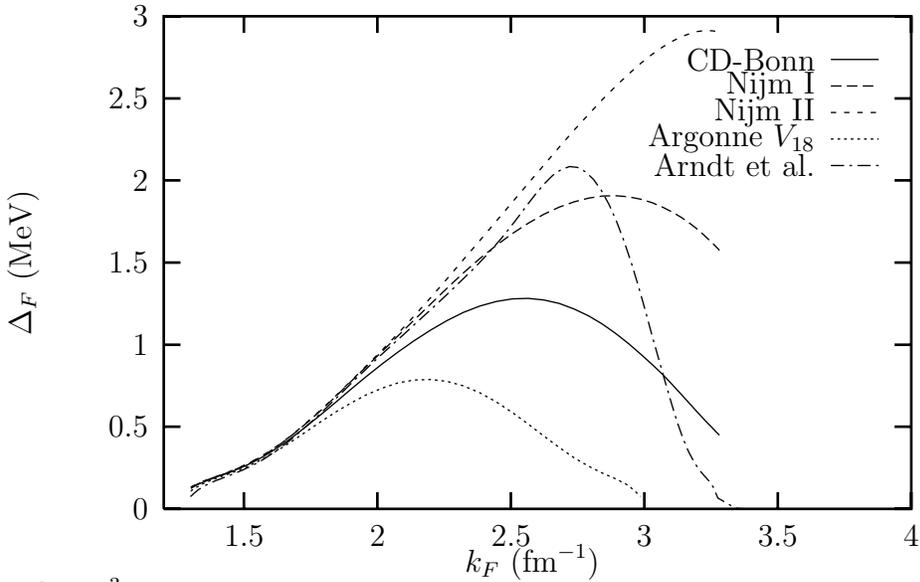
\begin{figure}
\setlength{\unitlength}{0.1bp}
\special{!
/gnudict 40 dict def
gnudict begin
/Color false def
/Solid false def
/gnulinewidth 5.000 def
/vshift -33 def
/dl {10 mul} def
/hpt 31.5 def
/vpt 31.5 def
/M {moveto} bind def
/L {lineto} bind def
/R {rmoveto} bind def
/V {rlineto} bind def
/vpt2 vpt 2 mul def
/hpt2 hpt 2 mul def
/Lshow { currentpoint stroke M
  0 vshift R show } def
/Rshow { currentpoint stroke M
  dup stringwidth pop neg vshift R show } def
/Cshow { currentpoint stroke M
  dup stringwidth pop -2 div vshift R show } def
/DL { Color {setrgbcolor Solid {pop []} if 0 setdash }
 {pop pop pop Solid {pop []} if 0 setdash} ifelse } def
/BL { stroke gnulinewidth 2 mul setlinewidth } def
/AL { stroke gnulinewidth 2 div setlinewidth } def
/PL { stroke gnulinewidth setlinewidth } def
/LTb { BL [] 0 0 0 DL } def
/LTa { AL [1 dl 2 dl] 0 setdash 0 0 0 setrgbcolor } def
/LT0 { PL [] 0 1 0 DL } def
/LT1 { PL [4 dl 2 dl] 0 0 1 DL } def
/LT2 { PL [2 dl 3 dl] 1 0 0 DL } def
/LT3 { PL [1 dl 1.5 dl] 1 0 1 DL } def
/LT4 { PL [5 dl 2 dl 1 dl 2 dl] 0 1 1 DL } def
/LT5 { PL [4 dl 3 dl 1 dl 3 dl] 1 1 0 DL } def
/LT6 { PL [2 dl 2 dl 2 dl 4 dl] 0 0 0 DL } def
/LT7 { PL [2 dl 2 dl 2 dl 2 dl 2 dl 4 dl] 1 0.3 0 DL } def
/LT8 { PL [2 dl 2 dl 2 dl 2 dl 2 dl 2 dl 2 dl 4 dl] 0.5 0.5 0.5 DL } def
/P { stroke [] 0 setdash
  currentlinewidth 2 div sub M
  0 currentlinewidth V stroke } def
/D { stroke [] 0 setdash 2 copy vpt add M
  hpt neg vpt neg V hpt vpt neg V
  hpt vpt V hpt neg vpt V closepath stroke
  P } def
/A { stroke [] 0 setdash vpt sub M 0 vpt2 V
  currentpoint stroke M
  hpt neg vpt neg R hpt2 0 V stroke
  } def
/B { stroke [] 0 setdash 2 copy exch hpt sub exch vpt add M
  0 vpt2 neg V hpt2 0 V 0 vpt2 V
  hpt2 neg 0 V closepath stroke
  P } def
/C { stroke [] 0 setdash exch hpt sub exch vpt add M
  hpt2 vpt2 neg V currentpoint stroke M
  hpt2 neg 0 R hpt2 vpt2 V stroke } def
/T { stroke [] 0 setdash 2 copy vpt 1.12 mul add M
  hpt neg vpt -1.62 mul V
  hpt 2 mul 0 V
  hpt neg vpt 1.62 mul V closepath stroke
  P  } def
/S { 2 copy A C} def
end
}
\begin{picture}(3600,2160)(0,0)
\special{"
gnudict begin
gsave
50 50 translate
0.100 0.100 scale
0 setgray
/Helvetica findfont 100 scalefont setfont
newpath
-500.000000 -500.000000 translate
LTa
600 251 M
2817 0 V
LTb
600 251 M
63 0 V
2754 0 R
-63 0 V
600 561 M
63 0 V
2754 0 R
-63 0 V
600 870 M
63 0 V
2754 0 R
-63 0 V
600 1180 M
63 0 V
2754 0 R
-63 0 V
600 1490 M
63 0 V
2754 0 R
-63 0 V
600 1799 M
63 0 V
2754 0 R
-63 0 V
600 2109 M
63 0 V
2754 0 R
-63 0 V
902 251 M
0 63 V
0 1795 R
0 -63 V
1405 251 M
0 63 V
0 1795 R
0 -63 V
1908 251 M
0 63 V
0 1795 R
0 -63 V
2411 251 M
0 63 V
0 1795 R
0 -63 V
2914 251 M
0 63 V
0 1795 R
0 -63 V
3417 251 M
0 63 V
0 1795 R
0 -63 V
600 251 M
2817 0 V
0 1858 V
-2817 0 V
600 251 L
LT0
3114 1946 M
180 0 V
702 328 M
5 3 V
8 5 V
13 7 V
16 7 V
20 8 V
23 9 V
27 10 V
31 10 V
33 14 V
37 16 V
40 20 V
43 25 V
46 30 V
48 35 V
50 39 V
53 43 V
54 45 V
56 48 V
58 48 V
59 47 V
60 45 V
61 43 V
62 40 V
62 36 V
62 31 V
62 24 V
61 17 V
61 10 V
60 2 V
59 -6 V
58 -13 V
56 -21 V
55 -26 V
52 -31 V
51 -36 V
48 -38 V
45 -39 V
43 -39 V
40 -38 V
37 -39 V
34 -37 V
30 -33 V
27 -30 V
23 -25 V
20 -21 V
17 -17 V
12 -13 V
9 -9 V
5 -5 V
LT1
3114 1846 M
180 0 V
702 331 M
5 3 V
8 4 V
13 7 V
16 7 V
20 8 V
23 9 V
27 9 V
31 11 V
33 14 V
37 17 V
40 21 V
43 26 V
46 32 V
48 37 V
50 42 V
53 47 V
54 51 V
56 55 V
58 56 V
59 57 V
60 59 V
61 60 V
62 60 V
62 60 V
62 58 V
62 54 V
61 49 V
61 45 V
60 39 V
59 34 V
58 27 V
56 20 V
55 14 V
52 8 V
51 1 V
48 -4 V
45 -9 V
43 -13 V
40 -17 V
37 -19 V
34 -20 V
30 -21 V
27 -21 V
23 -20 V
20 -19 V
17 -16 V
12 -12 V
9 -9 V
5 -6 V
LT2
3114 1746 M
180 0 V
702 333 M
5 3 V
8 4 V
13 6 V
16 8 V
20 7 V
23 9 V
27 9 V
31 12 V
33 13 V
37 17 V
40 21 V
43 26 V
46 32 V
48 37 V
50 43 V
53 47 V
54 51 V
56 55 V
58 57 V
59 59 V
60 63 V
61 66 V
62 69 V
62 71 V
62 73 V
62 72 V
61 72 V
61 72 V
60 71 V
59 69 V
58 68 V
56 64 V
55 60 V
52 56 V
51 51 V
48 45 V
45 40 V
43 34 V
40 28 V
37 22 V
34 17 V
30 11 V
27 8 V
23 3 V
20 1 V
17 -1 V
12 -2 V
9 -2 V
5 -1 V
LT3
3114 1646 M
180 0 V
702 318 M
5 4 V
8 6 V
13 7 V
16 9 V
20 9 V
23 9 V
27 10 V
31 11 V
33 12 V
37 16 V
40 19 V
43 24 V
46 27 V
48 31 V
50 34 V
53 36 V
54 35 V
56 33 V
58 31 V
59 25 V
60 20 V
61 11 V
62 2 V
62 -7 V
62 -17 V
62 -26 V
61 -34 V
61 -40 V
60 -44 V
59 -45 V
58 -43 V
56 -39 V
55 -32 V
52 -26 V
51 -22 V
48 -25 V
45 -41 V
LT4
3114 1546 M
180 0 V
701 297 M
1 2 V
3 2 V
3 3 V
4 4 V
6 5 V
6 5 V
7 6 V
9 5 V
9 6 V
11 6 V
11 5 V
12 6 V
14 6 V
14 5 V
15 6 V
16 6 V
17 7 V
18 7 V
19 8 V
19 9 V
21 10 V
21 11 V
22 13 V
22 15 V
24 16 V
24 18 V
25 19 V
25 21 V
26 23 V
27 24 V
28 25 V
28 27 V
28 27 V
29 28 V
29 29 V
30 30 V
31 29 V
30 29 V
31 29 V
32 29 V
32 29 V
32 29 V
32 30 V
32 29 V
33 30 V
33 31 V
33 31 V
32 31 V
33 32 V
33 32 V
33 35 V
33 36 V
33 38 V
33 41 V
33 43 V
32 43 V
32 41 V
32 38 V
32 32 V
32 22 V
31 11 V
30 -3 V
31 -17 V
30 -29 V
29 -41 V
29 -53 V
28 -63 V
28 -74 V
28 -83 V
27 -89 V
26 -92 V
25 -93 V
25 -88 V
24 -85 V
24 -82 V
22 -72 V
22 -62 V
21 -50 V
21 -39 V
19 -29 V
19 -22 V
18 -20 V
17 -25 V
16 -39 V
66 -38 V
11 -1 V
9 -1 V
8 0 V
8 -1 V
6 0 V
6 0 V
4 0 V
3 0 V
3 0 V
1 0 V
stroke
grestore
end
showpage
}
\put(3054,1546){\makebox(0,0)[r]{Arndt et al.}}
\put(3054,1646){\makebox(0,0)[r]{Argonne $V_{18}$}}
\put(3054,1746){\makebox(0,0)[r]{Nijm II}}
\put(3054,1846){\makebox(0,0)[r]{Nijm I}}
\put(3054,1946){\makebox(0,0)[r]{CD-Bonn}}
\put(2008,51){\makebox(0,0){$k_F$ (${\rm fm}^{-1}$) }}
\put(100,1180){%
\special{ps: gsave currentpoint currentpoint translate
270 rotate neg exch neg exch translate}%
\makebox(0,0)[b]{\shortstack{$\Delta_F$ (${\rm MeV}$) }}%
\special{ps: currentpoint grestore moveto}%
}
\put(3417,151){\makebox(0,0){4}}
\put(2914,151){\makebox(0,0){3.5}}
\put(2411,151){\makebox(0,0){3}}
\put(1908,151){\makebox(0,0){2.5}}
\put(1405,151){\makebox(0,0){2}}
\put(902,151){\makebox(0,0){1.5}}
\put(540,2109){\makebox(0,0)[r]{3}}
\put(540,1799){\makebox(0,0)[r]{2.5}}
\put(540,1490){\makebox(0,0)[r]{2}}
\put(540,1180){\makebox(0,0)[r]{1.5}}
\put(540,870){\makebox(0,0)[r]{1}}
\put(540,561){\makebox(0,0)[r]{0.5}}
\put(540,251){\makebox(0,0)[r]{0}}
\end{picture}
\caption{$^3P_2$ gap calculated with separable potentials  
         constructed directly from the $^3P_2$ phase shifts.}
\label{fig:pshiftgap}
\end{figure}

Using phase shifts from the analysis of Arndt et al.~\cite{arndt97,said98}, 
we constructed an interaction for the 
$^3P_2$ channel according to the prescription above, and then 
proceeded to solve Eq.~(\ref{eq:sepeq5}) for $\Delta_F$.  
The results are shown in Fig.~\ref{fig:pshiftgap}.
For comparison we also display the results of the following calculation 
for the various potentials:  
we took the phase shifts at energies 
up to 1100 MeV computed earlier and shown in Fig.~\ref{fig:phaseshift}. 
From these we constructed a rank-one separable approximation to the 
$^3P_2$ wave of the various potentials, as described above, and then 
used this to solve the gap equation.  As such, we have a as close as possible
link with the calculation based solely on the phase shifts of 
Arndt et al.~\cite{arndt97,said98}.
This allows us in turn to see directly the consequences of the  
failure of the potentials to fit the high-energy $^3P_2$ phase shifts. 
When looking at Fig.~\ref{fig:pshiftgap} and reading the 
following discussion, one should bear in mind that the gap 
has an exponential dependence on the interaction, so quite 
small differences in the matrix elements of the interaction 
can be translated into large differences in the energy gap.  
But this also makes the gap a good quantity to use when 
comparing interactions, as any difference is magnified. 

Although the approximation made here should not be taken too seriously, 
the results indicate some important conclusions about the 
$^3P_2$ waves of the recent nucleon-nucleon interactions.  
All seem to have about the right amount of attraction at 
densities below $k_F\approx 2.0\;{\rm fm}^{-1}$.  
Between $k_F\approx 2.0\;{\rm fm}^{-1}$ and $k_F\approx 3.0\;{\rm fm}^{-1}$ 
all interactions except Nijm-II seem to be a bit too repulsive.  
Above $k_F\approx 3.0$, 
Argonne $V_{18}$ is probably too repulsive, while Nijm I and II are 
most certainly too attractive, and the same probably also holds for 
the CD-Bonn.  If one uses the weak-coupling expression for the gap, 
\begin{equation}
  \Delta_F\approx 2 \epsilon_F e^{-V_{11}(k_F,k_F)/N(0)} \:, 
\label{eq:sepeq6}
\end{equation}
where $\epsilon_F$ is the Fermi energy and $N(0)$ the density of 
states at the Fermi level, one sees that the gap vanishes  
where the interaction goes to zero.  In our phase-shift approximation, 
this happens where the phase shifts change sign, at $k_F\approx 
3.6\;{\rm fm}^{-1}$.  The Argonne $V_{18}$ gap then seems to 
disappear somewhat too early, while the other potentials give gaps which 
exist up to what is probably unrealistically high densities.  

\section{Conclusion}
\label{sec:conclusions}
 
We have presented new calculations of the pairing gap in the 
$^3P_2$-$^3F_2$ channel for pure neutron matter as a function of density. 
With these calculations we have aimed at establishing  on a firm basis the 
numerical value of
the gap once the bare nucleon-nucleon interaction is used as the
pairing interaction, since in this context contradictory results
have been presented in the literature. 
Three different numerical methods to solve the pairing gap have been
employed in this paper> Since all three methods gave  
the same results, the pairing gaps we have obtained 
should be reliable from a technical point of view.    

\begin{figure}
\includegraphics[totalheight=17.5cm,angle=0,bb=0 80 350 730]{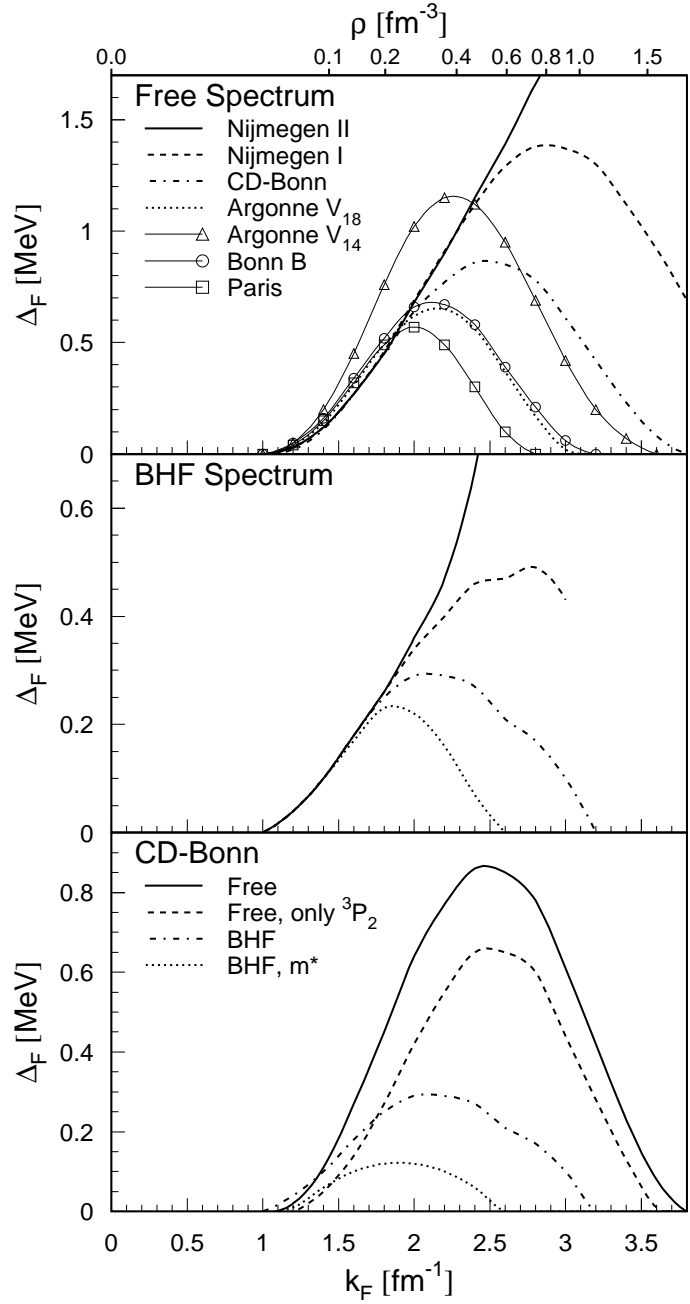}
\caption{Top panel: The angle-averaged $^3P_2$-$^3F_2$ gap in neutron matter
         depending on the Fermi momentum, evaluated with free 
         single-particle spectrum and different nucleon-nucleon potentials. 
         Central panel: The gap evaluated with BHF spectra.
         Bottom panel: The gap with the CD-Bonn potential in different 
         approximation schemes.}
\label{fig:gaps}
\end{figure}

However, our calculations have revealed that the behavior of 
the $^3P_2$-$^3F_2$ gap at densities above $k_F \approx 2.0\;{\rm fm}^{-1}$, 
corresponding to $\rho \approx 1.7\rho_0$, where $\rho_0$ 
is the nuclear matter saturation density, must be considered as 
largely unknown.  
Up to this point the gap is increasing
(the values at $k_F=2.0\;{\rm fm}^{-1}$ 
are about 0.6 MeV with free single-particle spectrum,
and about 0.3 MeV with BHF spectrum, independent of the potential),
but how far in density this increase continues, depends on the individual
potentials, in line with their extrapolations of the $^3P_2$ phase
shift predictions.
Bearing in mind that the Nijm-II potential fitted 
the empirical $^3P_2$ phase shift rather well up to 
$E_{\rm lab}\approx 600\;{\rm MeV}$ 
($k_F\approx 2.7\;{\rm fm}^{-1}$), 
we can deduce from Fig.~\ref{fig:gaps} that the maximum 
gap with a free spectrum is probably below $1\;{\rm MeV}$.  
How high up in density the gap exists must be left as an open question, 
although the phase shifts indicate that the gap should disappear 
at around $k_F=3.6\;{\rm fm}^{-1}$, corresponding to $\rho\approx 10\rho_0$.  
At this point also the purely nucleonic treatment of the dense medium
is surely inappropriate.

Before a precise calculation of the $^3P_2$-$^3F_2$ pairing gap can be made, 
one therefore needs a nucleon-nucleon potential that fits 
the phase shifts up to $E_{\rm lab}\approx 1\;{\rm GeV}$ accurately.  
To us, the construction of potential models in which the 
inelasticities above $E_{\rm lab}=350\;{\rm MeV}$ 
due to the opening of the ${\rm N}\Delta$ channel are taken 
into account,
seems to be more urgent than the evaluation of 
polarization effects on the $^3P_2$-$^3F_2$ gap with the existing potential models.   

\acknowledgements

We would like to thank John Clark, Umberto Lombardo and Eivind Osnes for 
interesting discussions.   
Thanks are also due to Ruprecht Machleidt for providing us 
with useful information about potentials and phase shifts 
in the $^3P_2$ wave.

\begin{table}[t]
\begin{center}
\begin{tabular}{llllllll} 
\multicolumn{1}{c}{$k_F\;({\rm fm}^{-1}$)}&\multicolumn{1}{c}{Bonn B}&
\multicolumn{1}{c}{Paris}&\multicolumn{1}{c}{$V_{14}$}& 
\multicolumn{1}{c}{CD-Bonn}&\multicolumn{1}{c}{$V_{18}$}&
\multicolumn{1}{c}{Nijm I}&\multicolumn{1}{c}{Nijm II} \\ \hline  
     1.2  & 0.05 & 0.04  & 0.05 & 0.03 & 0.04 & 0.03 & 0.03  \\
     1.4  & 0.16 & 0.15  & 0.19 & 0.11 & 0.14 & 0.12 & 0.12  \\
     1.6  & 0.35 & 0.32  & 0.45 & 0.27 & 0.31 & 0.27 & 0.27  \\
     1.8  & 0.52 & 0.49  & 0.75 & 0.45 & 0.49 & 0.47 & 0.45 \\
     2.0  & 0.66 & 0.57  & 1.02 & 0.64 & 0.62 & 0.69 & 0.68 \\
     2.2  & 0.67 & 0.49  & 1.14 & 0.77 & 0.65 & 0.91 & 0.90 \\
     2.4  & 0.58 & 0.30  & 1.13 & 0.86 & 0.56 & 1.12 & 1.15 \\
     2.6  & 0.39 & 0.10  & 0.95 & 0.85 & 0.37 & 1.26 & 1.39 \\
     2.8  & 0.21 & ---  & 0.70 & 0.78 & 0.17 & 1.38 & 1.66 \\
     3.0  & 0.06 & ---  & --- & 0.61 & 0.02 & 1.37 & 1.90 \\ \hline  
\end{tabular}
\caption{Collection of $^3P_2$-$^3F_2$ energy gaps (in MeV) for the 
various potentials considered in this paper.  
Free single-particle energies have been used.}
\label{tab:tab1}
\end{center}
\end{table}

\begin{table}[t]
\begin{center}
\begin{tabular}{llllllll} 
\multicolumn{1}{c}{$k_F\;({\rm fm}^{-1})$}&\multicolumn{1}{c}{Bonn B}&
\multicolumn{1}{c}{Paris}&\multicolumn{1}{c}{$V_{14}$}& 
\multicolumn{1}{c}{CD-Bonn}&\multicolumn{1}{c}{$V_{18}$}&
\multicolumn{1}{c}{Nijm I}&\multicolumn{1}{c}{Nijm II} \\ \hline  
     1.2  & 0.05 & 0.04  & 0.05 & 0.04 & 0.04  & 0.04  & 0.04  \\
     1.4  & 0.16 & 0.11  & 0.18 & 0.10 & 0.10  & 0.10  & 0.10  \\
     1.6  & 0.34 & 0.22  & 0.38 & 0.18 & 0.17  & 0.18  & 0.18  \\
     1.8  & 0.52 & 0.26  & 0.60 & 0.25 & 0.23  & 0.26  & 0.26  \\
     2.0  & 0.64 & 0.22  & 0.74 & 0.29 & 0.22  & 0.34  & 0.36  \\
     2.2  & 0.65 & 0.14  & 0.75 & 0.29 & 0.16  & 0.40  & 0.47  \\
     2.4  & 0.56 & 0.01  & 0.66 & 0.27 & 0.07  & 0.46  & 0.67  \\
     2.6  & 0.37 & ---  & 0.42 & 0.21 & ---        & 0.47  & 0.99  \\
     2.8  & 0.19 & ---  & 0.23 & 0.17 & ---        & 0.49  & 1.74  \\
     3.0  & 0.02 & ---  & 0.08 & 0.11 & ---        & 0.43  & 3.14  \\ \hline
\end{tabular}
\caption{Collection of $^3P_2$-$^3F_2$ energy gaps (in MeV) for the 
various potentials considered in this paper.  
BHF single-particle energies have been used.}
\label{tab:tab2}
\end{center}
\end{table}

\end{document}